# Joint analysis of Rayleigh-wave dispersion curves and diffuse-field HVSR for site characterization: The case of El Ejido town (SE Spain)

by

Antonio García-Jerez[a], Helena Seivane[a], Manuel Navarro[a], Marcos Martínez-Segura[b], José Piña-Flores[c]

[a] Department of Chemistry and Physics, University of Almeria, 04120 Almeria, Spain.

[b] Department of Mining, Geological and Cartographic Engineering. Polytechnic University of Cartagena, 30203 Cartagena, Murcia, Spain.

[c] Institute of Engineering, National Autonomous University of Mexico, Mexico.

Corresponding author: A.G.-J. Department of Chemistry and Physics, University of Almeria, 04120 Almeria, Spain. Email: agarcia-jerez@ual.es.

**Abstract**

The location of El Ejido town over a deep sedimentary basin in a zone of high seismicity in the Spanish context has motivated research on its seismic response characterization. To this aim, S-wave velocity models have been obtained from joint inversion of Rayleigh wave dispersion curves and full-wavefield modelling of the horizontal-to-vertical spectral ratio of ambient noise (HVSR) under the diffuse field assumption (DFA). Combination of spatial autocorrelation surveys (SPAC) with array apertures of several hundred metres and HVSRs displaying low-frequency peaks allowed to characterize deep ground features down to the Triassic bedrock. Predominant periods in the town ranged from 0.8 to 2.3 s, growing towards the SE, with few secondary peaks at higher frequencies.

The shallow structure has been explored by means of geotechnical surveys, Multichannel Analysis of Surface Waves (MASW) and SPAC analysis in small-aperture arrays. Resulting models support a general classification of the ground as stiff soil.

1. Introduction

As cities grow up, the number of elements exposed to risk becomes bigger and updated strategies to reduce the seismic risk are required. This is the case of El Ejido, a new town in southeast Spain with a constant growth rate which has trebled its population since the early eighties. It is located 10 kilometres from the Alboran sea coast, the most seismically active region in the country. The significant seismic hazard in this area is supported by historical seismic reports [1] and has been evidenced by recent strong earthquake swarms occurred in Alboran Sea [2]. This perspective made it worthwhile for us to carry out a seismic microzonation in terms of seismic velocities and predominant periods of the ground motion.

Regarding to the local ground conditions of El Ejido, there are two remarkable features which turn this area into a singular framework. It is located atop Campo de Dalías, which is a deep sedimentary basin capable of developing low-frequency resonances of seismic waves, whereas the relatively hard soils found near surface could lead to underestimating the seismic hazard that really exists. European seismic code [3] includes a soil characterization in terms of shear wave velocities. Nevertheless, such classifications of soil based solely on the average shear-wave velocities for the upper 30 m ($V_{S30}$) ignore velocity contrasts between soil layers below 30 m, especially in cases of deep sites as Campo de Dalías [4–6] which could result in a less accurate determination of elastic response spectra. Moreover, the ratio between the resonant periods and the natural period of the building structures has been demonstrated as an explanatory parameter of damage increment in affected buildings [7].

Modern structural design codes [8,9] begin to include the site period in the definition of soil classes and researchers are proposing new site classifications based on this property [10,11]. It is a well-known fact that unconsolidated deposits amplify seismic waves [12]. Destructive earthquakes as the 1985 Michoacán and 2015 Kathmandu events, as well as moderate ones like the 1993-1994 Adra earthquakes [13] are evidence of this. On the other hand, long-period ground motion is increasingly often considered as an essential point due to its interaction with large-scale structures [14,15] because long predominant periods linked to deep basins might play a role as important as softer sediments in shallower basins.

Geological and geotechnical information gives evidence of the actual landform conditions for shallow and deep layers through a characterization of existing materials, being the shear-wave velocity ($V_S$) structure the most relevant parameter for site effect estimation. A long-established parameter in geotechnical studies, the $N_{SPT}$ value is widely used for estimation of $V_S$ ground structure where more specific methods are not available (e.g. [16–18]).

The $V_S$ structure of ground materials can be also obtained by processing short-period Rayleigh waves by means of techniques that take advantage of their dispersive properties when they travel through a layered media. The phase-velocity $c_R(f)$ is mainly determined by the $V_S$ structure with slight dependence on the P-wave velocity ($V_P$) and density ($\rho$). Surface waves provide the highest possible signal-to-noise ratio for shallow seismic exploration. Consequently, the field works for data acquisition and subsequent data analysis become cost-effective in comparison with other classical techniques, ensuring reliable results [19–21].

The spatial autocorrelation method (SPAC, [22]) and the horizontal-to-vertical spectral ratio (HVSR) provide a theoretical basis for ground characterization from analysis of seismic noise signals. The HVSR technique [23] is nowadays used as the standard tool for obtaining the fundamental site period meanwhile array-based techniques provide elastic parameters variations with depth [22]. The combination of these two methods in joint inversion schemes allows resolving lower frequency bands and the effective identification of deeper ground interfaces (e.g. [24–28]). The diffuse wavefield approach (DFA, [29]), based on the seismic interferometry theory, provides a suitable theoretical framework for modelling these observations.

This work is focused on the microzonation of El Ejido through available geological and geotechnical information, active-source surface-wave methods and ambient seismic noise surveys. To this aim, geological and geotechnical data have been compiled and described. Application of the MASW method along six profiles enhanced the resolution in the upper layers. The possibility of estimating $V_{S30}$ from the phase velocity of Rayleigh waves of a particular wavelength, $c_\lambda$, has been evaluated using the MASW dataset. Then, a dense seismic microzonation based on the HVSR technique has been carried out. HVSRs have been used for inversion of local 1-D velocity profiles, combined with geological information and/or dispersion curves retrieved from array techniques, and 2D cross-sections have been built from several local models. An algorithm based on the DFA [30] was employed for this task. Instrumental choices and limitations of techniques used for signal analysis were taken into account to interpret the results accurately.

This study is the first stage for future seismic risk mitigation plans which should consider local seismic response effects, based on the whole set of soil dynamic parameters, to assess the seismic hazard level of the populated towns in this coastal plain.

## 2. Geological and Geotechnical settings

The town of El Ejido is located in Campo de Dalías, a basin bordered by the Alboran Sea to the south and by Sierra de Gador mountain range to the north (Fig. 1). From a tectonic point of view, Campo de Dalías basin is located in the Internal Zone of the Alpine Betic Chain, being an emerged portion of the Alboran Sea [31]. It has an extension of around 330 km$^2$ spanning 33 km long (in E-W direction) and 14 km maximum wide (in N-S direction). El Ejido is over thick sedimentary fillings with ages between Mid-Miocene and Holocene as concluded in previous general studies of Campo de Dalías basin (e.g. [32,33]). These sediments overlay a folded Palaeozoic-Triassic metamorphic complex, denoted as Alpujárride complex, which is settled as the bedrock of the coastal plain. Sierra de Gádor is the most representative outcrop of this complex, that remains as the bedrock under the marine sediments in Alboran Sea [34] (Fig. 2).

The basin basement is composed of dolomites and limestones slightly metamorphosed in the upper part, although phyllites and quartzs can be found on the occidental and oriental borders of Sierra de Gádor [33]. Miocene sediments are deposited discordantly over this folded and faulted basement, including rocks of marly, clayey and sandy-silty nature in the study area. The Miocene succession is overlaid, also discordantly, by a Pliocene sequence mainly composed of calcarenites, silts, marls, sands and conglomerates [32,35]. Quaternary deposits lay on top of the upper Pliocene and can be either marine or continental in origin [33]. Those of continental origin are made up of red silts and they are abundantly found on the surface of El Ejido as well as marine terraces (Fig. 2a). Pliocene calcarenites are also found in a small percentage on El Ejido surface, but their bigger outcrop is found in the southwestern sector of Campo Dalías plain. Therefore, Quaternary deposits have a variable lateral continuity while Pliocene unit has a higher continuity on the scale of the study area [35] (Fig. 2b).

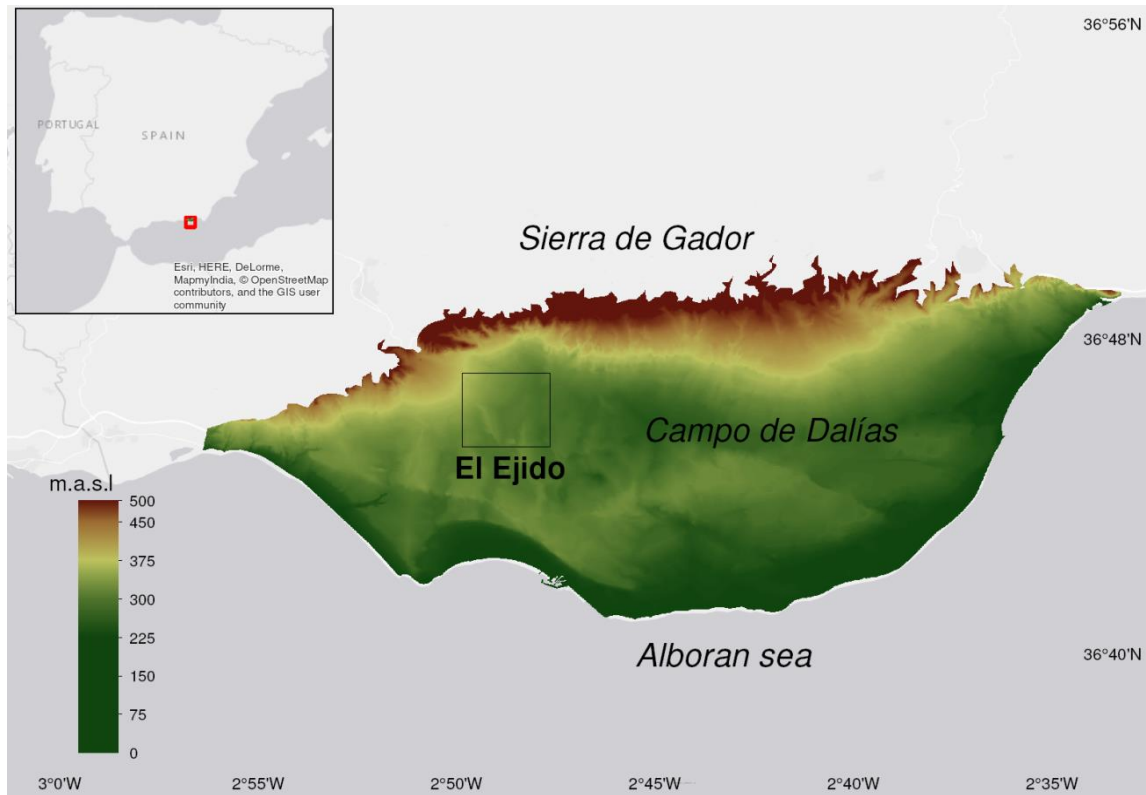

**Fig. 1.** Location of the study area. Background colour of Campo de Dalías displays the digital elevation model.

Industrial intensive agriculture is the main economic activity in Campo de Dalías using a large percentage of the coastal plain lands. Such activities find on underground aquifers their main water supply, resulting in a huge number of dug wells throughout this area. These resources are useful for interpretation of the passive seismic surveys and they enabled us to draw various preliminary cross sections of the sedimentary cover near El Ejido (Fig. 2b). Borehole data along with some previous studies based on seismic surveys [36] provide us with *a priori* information of average seismic velocities and densities. Materials of the Pliocene unit have been attributed with mean seismic velocities ($V_P$) around 2.6 km/s [33]. Upper Tortonian calcarenites have estimated seismic velocities ($V_P$) around 3 km/s. On the other hand, 1-D velocity models in the Alboran Sea obtained by Grevemeyer et al. [37] attribute $V_P$ above 5 km/s to the first crustal layer, which would comprise the Alpujarride complex.

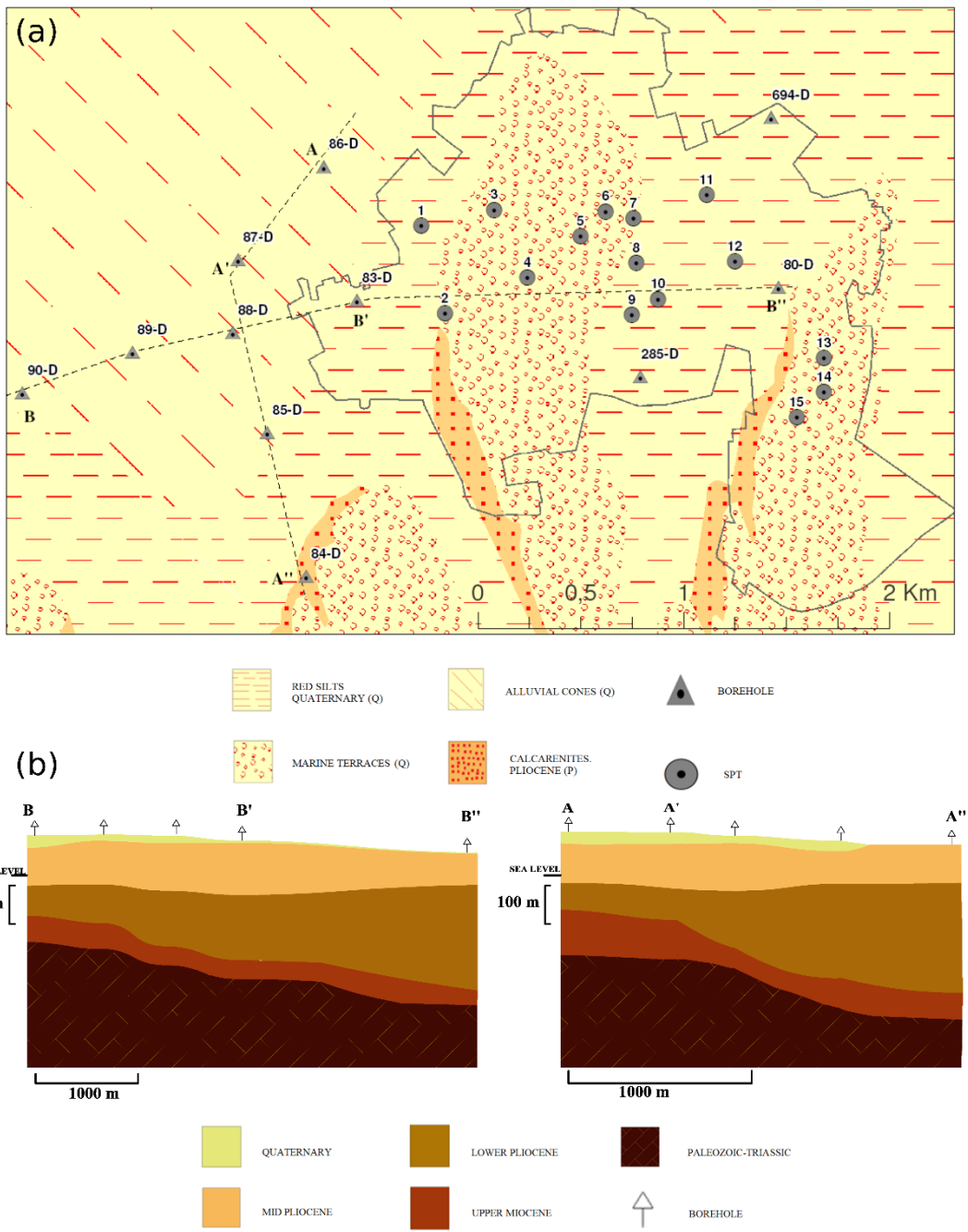

**Fig. 2.** a) Surface geological map of El Ejido area (modified from MAGNA, Spanish Geological Survey). The urban perimeter is denoted by the black line. Borehole locations are represented by triangles and the Standard Penetration Tests by circles b) Cross-sections along profiles sketched from available boreholes.

Geotechnical information extracted from 15 standard penetration tests (SPT) made for foundation design (Figure 3) has been used to investigate the composition and properties of the shallow ground layers. Quaternary red silts and clays cover the central flatland in El Ejido town, which is wedged in between two gently hills with N-S trend enclosed in the marine terraces area (Fig. 2a). The deepest geotechnical surveys show that these Quaternary deposits have thicknesses of up to 25 m in the central-southern zones (surveys 8, 10, 12) increasing up to 40 m in the NE part (694-D). They overlay Pliocene calcarenites which can be effectively considered as the geotechnical basement. Surveys performed on the hills show abundant gravelly levels (1, 5, 6), calcareous crusts (3, 4, 14) and conglomeratic rocks (13, 15) in the upper meters. Pliocene calcarenites are almost outcropping in surveys 2 and 80-D and at a depth of 9 m in survey 15. The deepest SPTs penetrated 15 m into this calcarenitic layer without reaching its bottom. Deep wells performed in the town (80-D, 285-D, 694-D) reached the bottom of the calcarenitic layer at depths between 60 and 90 m.

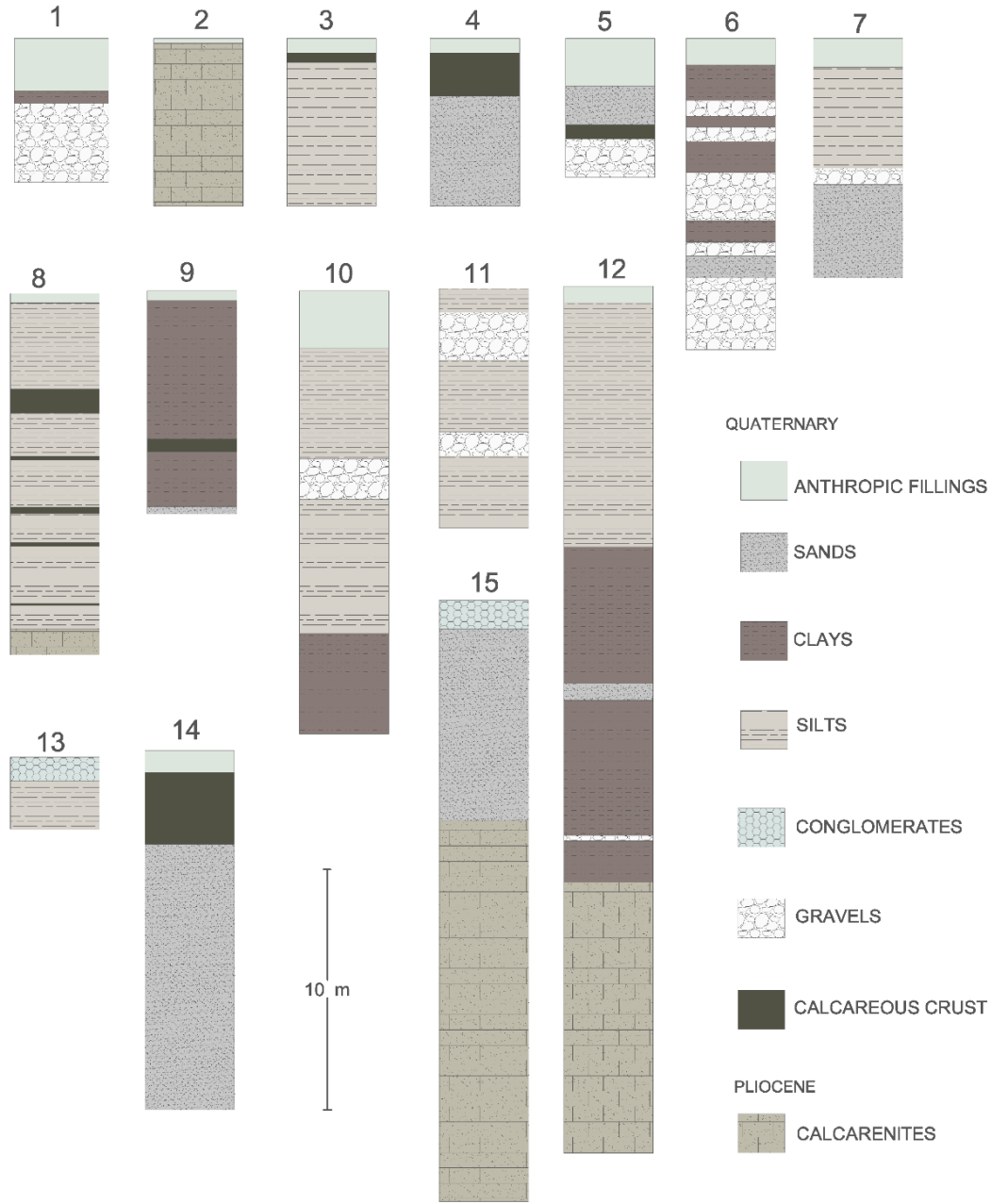

**Fig. 3.** Lithostratigraphic columns from studied geotechnical loggings. Locations of these logs can be found in Fig. 2a.

## 3. Analysis and results

A brief description of applied methodologies, data analysis and results obtained for determination of the shallow and deep ground structure in El Ejido town are presented below. The shallow $V_S$ structure has been estimated from geotechnical information and calculated using MASW and SPAC techniques. The deep ground structure is determined by joint inversion of HVSRs and dispersion curves retrieved from ambient noise records.

*3.1. Geotechnical tests*

As a first approach to estimate $V_S$ values, empirical relationships between this property and the SPT blow count ($N_{SPT}$) have been used [16–18,38]. Except for one case of 50 metres drilling, all the tests were performed very shallow, between 3 and 30 metres deep. The water table was found in the deepest borehole at 34 m in Pliocene calcarenites. Due to the lack of local $V_S$-$N_{SPT}$ relationships for the soils of the study area, the values obtained with several empirical formulae (Table 1) must be taken as a rough approximation.

**Table 1.** Empirical correlations based on $N_{SPT}$ vs $V_S$ to process the geotechnical tests

| Researcher(s) | Soil type | $V_S$ based on $N_{SPT}$-value (m/s) | $r^2$ |
|---|---|---|---|
| Ohta and Goto (1978) | Clays and silts | $85.6\, N^{0.34}$ | 0.73 |
| | Sand and gravels | $85.6 \cdot 1.048\, N^{0.34}$ | 0.73 |
| | Gravels | $85.6 \cdot 1.222\, N^{0.34}$ | 0.73 |
| Akin et al. (2011) | All alluvial Soils | $59.44\, N^{0.109}\, z^{0.426}$ | 0.79 |
| | All Pliocene Soils | $121.85\, N^{0.101}\, z^{0.216}$ | 0.88 |
| Tan et al. (2013) | Soft Soil | $101.34\, N^{0.2364}$ | 0.83 |
| | Stiff soil | $128.71\, N^{0.2833}$ | 0.65 |
| | Hard Soil and Soft Rock | $128.05\, N^{0.4081}$ | 0.73 |
| Yoshida and Motonori (1988) | 50% Gravel Soil | $60\, N^{0.25}\, \sigma'_{v0}{}^{0.14}$ | - |

$z$ = depth in metres, N = number of blow counts, $r^2$ = multiple correlation coefficient, $\sigma'_{v0}$=Vertical pressure in kPa.

Geotechnical data show a Quaternary unit varying widely in composition and interleaved with calcareous and conglomeratic crusts which lead to meet refusal of the in-situ tests even in the top. Stratigraphic sections generally have a first unit of artificial ground with an average thickness of 80 cm. Most of them drilled the Quaternary unit only. The vast majority of empirical relationships between $N_{SPT}$ and $V_S$ found in literature have been calculated for Quaternary soils while older materials are scarcely studied. Therefore, the possibility of comparison of estimated $V_S$ values for Pliocene calcarenites is very limited (Table 2).

According to Table 2, materials found on *in situ* tests were estimated to have average $V_S$ of 263 m/s for sands, 268 m/s for silts, 347 m/s for gravels and 297 m/s for clays. Despite the lack of local and specific correlations for calcarenites a $V_S$ mean value of 511 m/s has been are estimated for these rocks.

**Table 2.** Description of depths (minimum and maximum), N-value, real density ($\rho$) and $V_S$ computed from the $N_{SPT}$ value by using empirical relationships

| Soil Type | Depth Levels (min. – max.) (m) | N-value | | | $\rho$ (g/cm³) | | | $V_S$ (Ohta & Goto) (m/s) | | $V_S$ (Akin et. al) (m/s) | | $V_S$ (Tan et. al) (m/s) | | $V_S$ Yoshida and Motonori (1988) 50% gravels | |
|---|---|---|---|---|---|---|---|---|---|---|---|---|---|---|---|
| | | n | X | σ | n | X | σ | X | σ | X | σ | X | σ | X | σ |
| *Quaternary* | | | | | | | | | | | | | | | |
| Sands with gravels | 1.0-9.0 | 1 | 25 | - | 2 | 1.90 | 0.14 | 268 | - | 205 | - | 320 | - | | |
| Silts | 1.5-13.6 | 29 | 22 | 16 | 2 | 1.84 | 0.01 | 256 | 75 | 230 | 125 | 319 | 77 | | |
| Gravels | 2.3-8.7 | 18 | 46 | 18 | 1 | 1.72 | - | 379 | 43 | | | 376 | 35 | 287 | 34 |
| Clays | 1.9-24.8 | 26 | 33 | 18 | 1 | 1.94 | - | 282 | 64 | 264 | 86 | 346 | 67 | | |
| *Pliocene* | | | | | | | | | | | | | | | |
| Calcarenites | 3.6-47.4 | 19 | 49 | 22 | 2 | 2.15 | 0.07 | - | - | - | - | 666 | 145 | | |

n= number of data, X= mean value, σ = standard deviation.

*3.2. Multichannel Analysis of Surface Waves*

In November 2017, 3.4 km of linear transects arranged in six profiles were laid out through several streets of El Ejido (Fig. 4) for an active-source MASW survey. Locations of these transects were chosen to sample the main geological units described in Fig. 2. A Wacker Neuson BS60-4s vibratory rammer was used as seismic source (Fig. 5b). To reach high efficiency in terms of surveyed length per day, a towed land-streamer was built by using a heavy-duty fire hose (Fig. 5a), which enabled collecting data in a roll-along mode. The seismic equipment consisted of 24 geophones of 4.5 Hz natural frequency screwed onto metal plates with 2 m spacing and a SUMMIT II Compact recording unit (Fig. 5a). The offset from the seismic source to the first geophone was 4 metres. This linear array of 46 m in length was displaced 10 metres between consecutive shots.

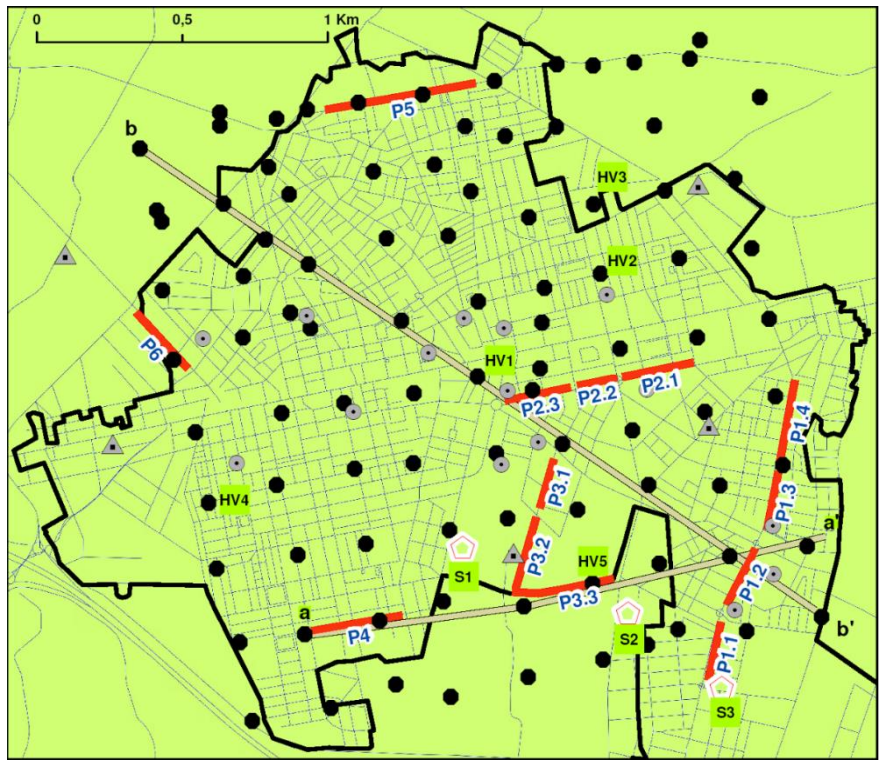

**Fig. 4.** Distribution of conducted seismic surveys and accessed boreholes in El Ejido urban area. Red lines represent MASW profiles. Black dots are HVSR measurement points (Section 3.3). Red pentagons illustrate the deployment points (S1, S2 and S3) of portable seismic arrays (Section 3.4). Grey circles and triangles show the locations of the analysed SPTs and boreholes (see details in Figs. 2 and 3). Lines a-a' and b-b' denote the profiles for which HVSR curves were inverted (Section 3.3.1).

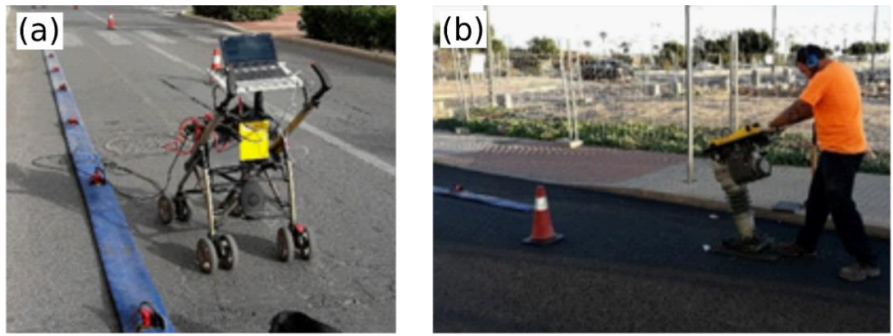

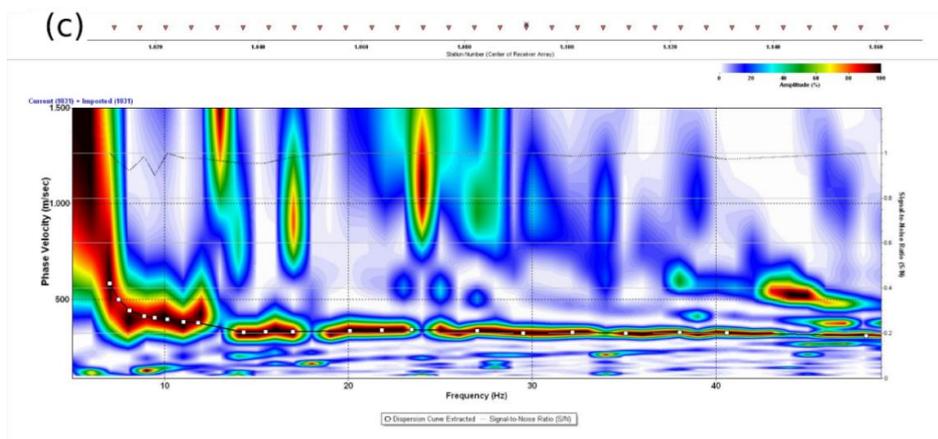

**Fig. 5.** Experimental setup for data acquisition of MASW data and example of dispersion curve retrieval. (a) Land-streamer of 46 m in length with 4.5 Hz geophones; (b) manual vibratory rammer as seismic source; (c) Dispersion curve picking for a shot gather belonging to profile P 3.3.

The seismic data were processed by using the "SurfSeis" software package from Kansas Geological Survey, USA. Dispersion curves were manually picked from each shot gather after applying the wavefield transformation into the frequency-velocity domain described by Park [39] (see also [20, 40]). This processing provided a set of dispersion curve fragments at relatively high frequencies, all of them comprised in the band 7 - 50 Hz. As an example of data processing, Fig. 5c depicts a dispersion diagram in which the fundamental mode is clearly observed. Since variations in the dispersion curves corresponding to individual profiles remained moderate, the respective average curves are shown for reference (Figure 6). For each profile, the dense series of dispersion curves were identified as fundamental-mode Rayleigh waves and separately inverted using the gradient-based iterative method described in Xia et al. [41]. The local 1D $V_S$ models were parametrized using 10 layers of variable thicknesses and unbounded seismic velocities $V_S$ and $V_P$, maintaining a constant Poisson's ratio of 0.25. The models presenting the minimum RMS error between experimental and theoretical dispersion curves were selected, allowing a maximum RMS of 0.5. Then, a 2D (horizontal position versus depth) $V_S$ model was built by means of kriging interpolation, assigning every 1D model to the midpoint of each receiver spread [40]. The resulting cross-sections are shown in Figure 7.

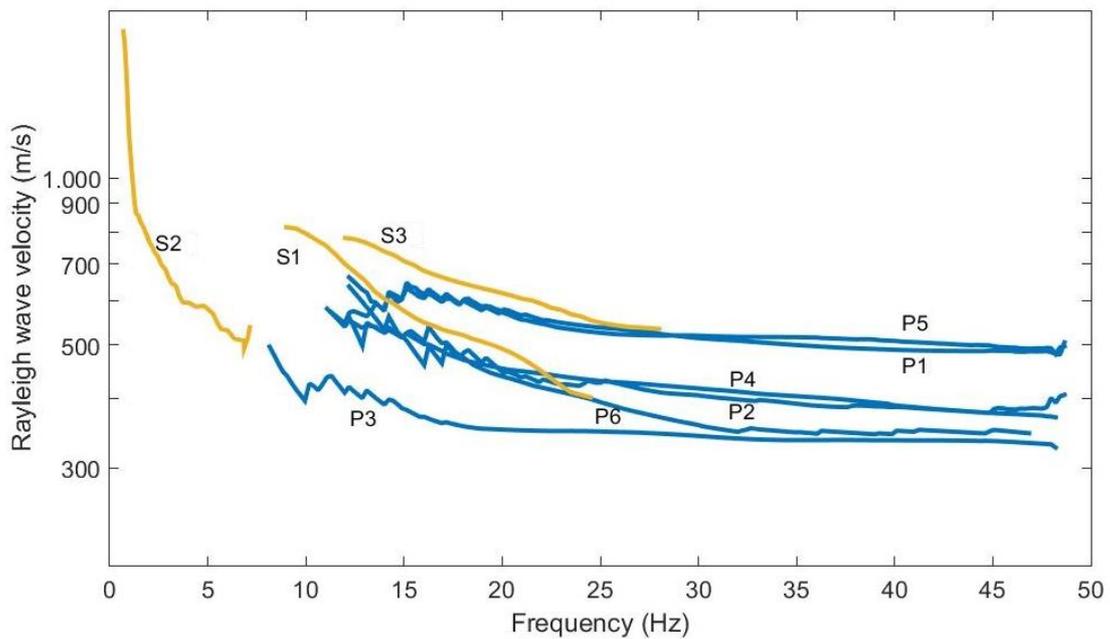

**Fig. 6.** Mean dispersion curves along the MASW profiles (Pn labels, blue colour). Dispersion curves computed from passive array experiments (Section 3.4) are shown with yellow lines and labelled as Sn. The respective locations can be found in Fig. 4.

It is usually considered that the maximum reliable Rayleigh wavelength for MASW surveying is of the order of the array length (e.g. [42]). Thus, approximate relationships lead to an effective surveying depth between 15 – 23 m for the here-employed array setup.

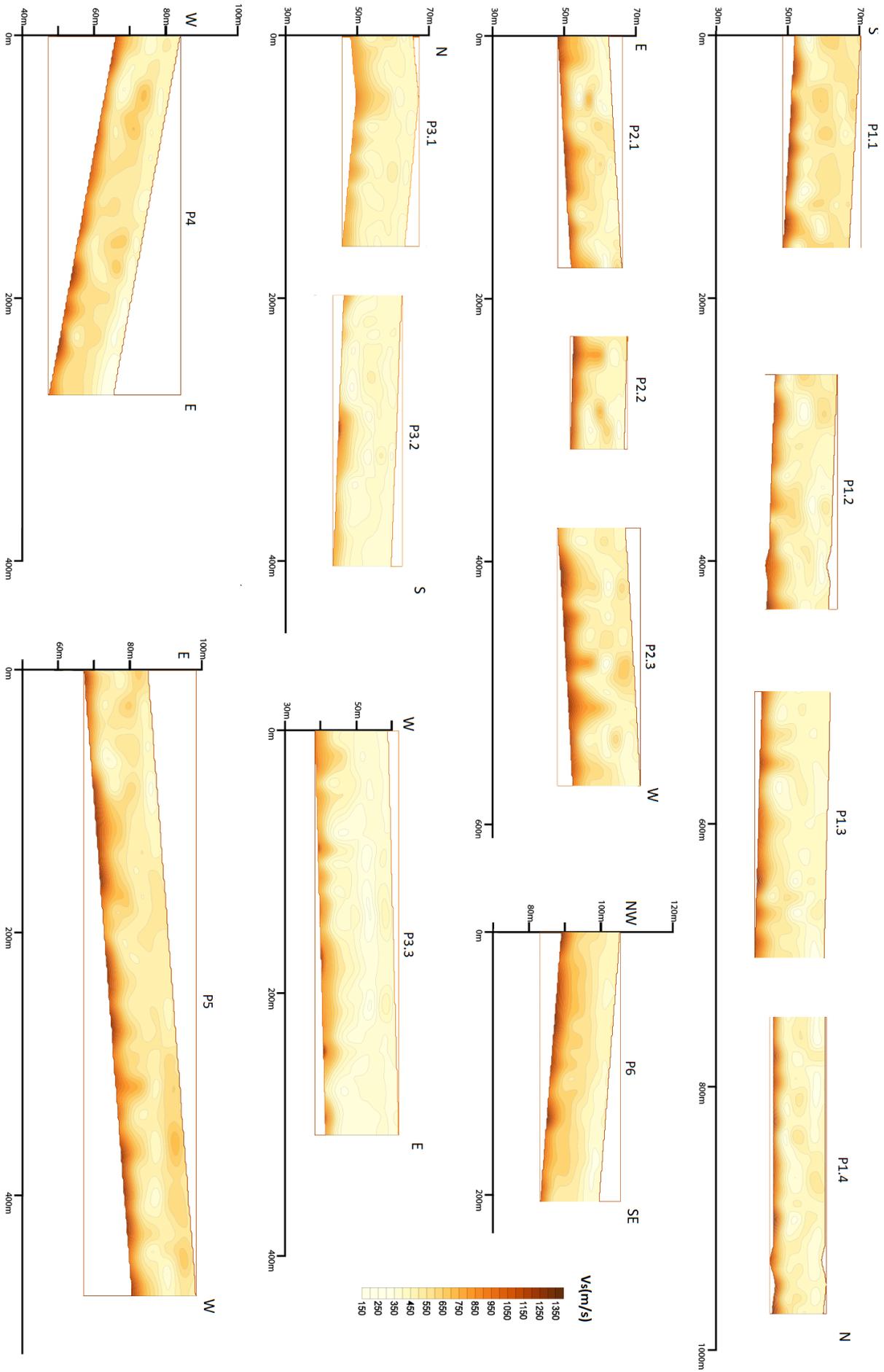

**Fig. 7.** MASW cross-sections obtained in El Ejido town. Profile locations are shown in Fig. 4.

Since each MASW line was designed to sample a particular geological unit, the $V_S$ structure shows a reasonable lateral uniformity along each cross-section. It allows synthesizing the results in terms of average velocities. Table 3 shows $V_{S30}$ values averaged along the profiles. Since the theoretical maximum penetration of the MASW method for our experimental setting is about 23 m, $V_{S30}$ estimates involve some extrapolation of the $V_S(z)$ structure, which has been performed assuming continuity of the bottom layer properties down to 30 m.

The highest average $V_{S30}$ values, $738 \pm 60$ and $725 \pm 70$ m/s, have been found in zones of marine terraces surveyed by profiles P1 and P5 (eastern hill and northern part of the town, respectively, Fig. 4). Velocities grow towards the north in P1 and they are slightly lower in the eastern part of P5, near a dry watercourse. Intermediate velocities between $605 \pm 40$ and $658 \pm 70$ m/s in $V_{S30}$ have been found in profiles P2, P4 and P6. Even though the shallow geology at P2 and P4 correspond to Quaternary red silts, presence of hard clays, cemented limes and calcareous crusts from 5 to 15 m deep is behind these relatively high velocities. Pliocene calcarenites were reached at depths of 14 and 25 m in the west and east ends of P2, respectively. The lowest $V_{S30}$ ($506 \pm 70$ m/s) corresponds to profile P3, in the central-southern part of the town, with decreasing values to the south. Nearby geotechnical logs (points 9 and 10 in Fig. 2a) show somewhat softer columns without significant cemented layers or having drilled calcarenites in the upper 19 m.

**Table 3.** Average $V_{S30}$ values and phase velocities for 40 and 45 m wavelength ($c_\lambda$) for each MASW profile. N shows the number of measurements considered

| Profile | N | $V_{S30}$ (m/s) | N | $c_{40}$ (m/s) | N | $c_{45}$ (m/s) |
|---------|---|---|---|---|---|---|
| P 1 | 75 | $738 \pm 60$ | 59 | $644 \pm 60$ | 30 | $684 \pm 50$ |
| P 2 | 44 | $658 \pm 70$ | 15 | $625 \pm 70$ | 8 | $640 \pm 80$ |
| P 3 | 67 | $506 \pm 70$ | 42 | $465 \pm 60$ | 32 | $494 \pm 70$ |
| P 4 | 27 | $605 \pm 40$ | 18 | $555 \pm 40$ | 10 | $570 \pm 40$ |
| P 5 | 47 | $725 \pm 70$ | 34 | $657 \pm 70$ | 21 | $672 \pm 80$ |
| P 6 | 21 | $606 \pm 50$ | 11 | $583 \pm 60$ | 6 | $612 \pm 80$ |

*3.3. HVSR technique*

Predominant period of soil was obtained in this work through seismic noise measurements by application of horizontal to vertical spectra ratio method (HVSR, [23]). This method is nowadays used as a standard tool for obtaining resonance frequencies of sediments when a high impedance contrast exists [43]. Over the last years, methods for modelling HVSR(*f*) were developed to be used in geophysical exploration. A recent review of these techniques has been made by Lunedei an Malischewsky [44].

In view of the relatively high sedimentary thicknesses (Fig. 2) and the expectation of fundamental frequencies distributed in a wide band, especially below 1 Hz, broadband seismometers were chosen to perform the seismic noise observations. Portable Güralp CMG-6TD triaxial seismometers were used to record the three components of ambient noise vibrations at 94 measurement points in the town (Fig. 4). Flat response in velocity of these seismometers is between 0.03 and 100 Hz. Every single station was equipped with a GPS and the measurements were sampled with a rate of 100 samples per second (sps). For each site, data acquisition lasted between 20 and 30 min depending on the level of local disturbances (pedestrians, nearby traffic, etc.). "Geopsy" software (http://www.geopsy.org; [45]) was used to compute HVSRs in the frequency range 0.25 - 15 Hz, employing 40 s long time windows overlapped 50%. Removal of those windows clearly contaminated with transients was done manually. Fourier amplitude spectra were smoothed through a Konno & Ohmachi window [46] with a bandwidth coefficient of 40.

Figure 8 shows the HVSRs obtained at four different sites which exemplify the typologies found: clear peak, broad peak, plateau and double peak. A univocal definition of the fundamental frequency, out of the total number of 94 measurement points, was obtained for 26 points included in the group of clear peaks (Fig. 8a).

Clarity of HVSR peaks obtained in El Ejido was tested following the recommendations proposed in SESAME guidelines, assessing criteria related to amplitude and frequency-dependent thresholds [47]. Only one of the HVSR curves was found with a weak peak, the rest were included in the four groups of typologies defined previously: 40 belong to broad-peak type (42% of our measurements), 27 were classified as clear peaks (28 %), 16 double peaks (18%) and 10 plateau (12%). Some trends can be observed in the spatial distribution of HVSR shape (Fig. 9). The northern part of the town is clearly represented by HVSR with broad peaks. In the central part, there is an irregular distribution including all the shape types. A majority of clear peaks are concentrated in the southeast part of the town. This distribution could be interpreted in terms of local complex stratigraphy as well as the effects of dipping interfaces near the basin edge [48–52].

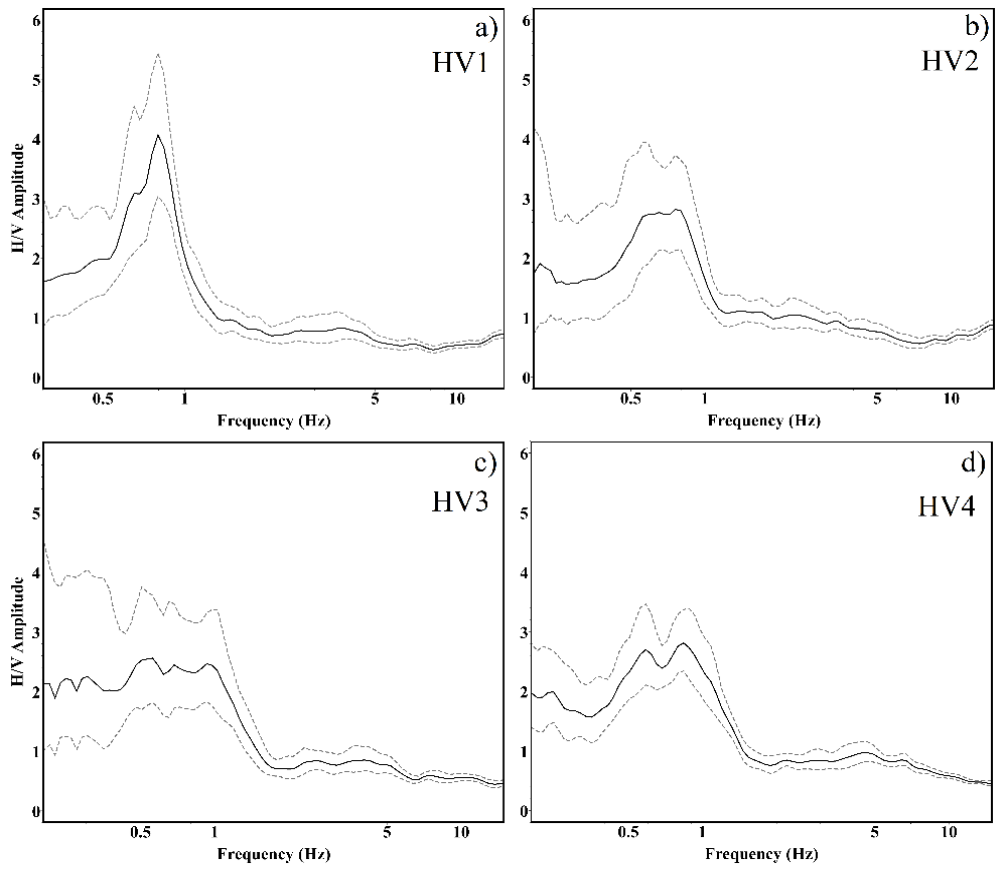

**Fig. 8.** Four instances of HVSR typologies found in El Ejido: a) curves with a single clear peak b) broad peaks whose widths vary from 0.2 to 0.4 Hz c) plateau-type shape with low amplitude and d) peaks showing narrower peaks on top.

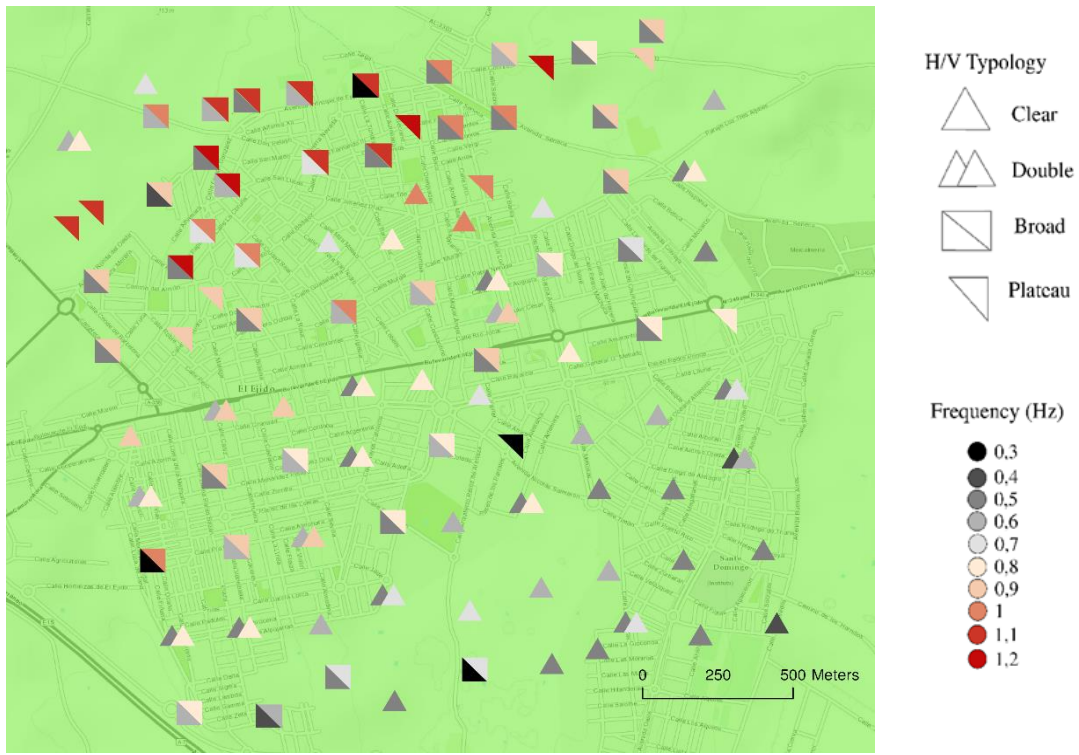

**Fig. 9.** Spatial distribution of HVSR peak frequencies in El Ejido urban area. Broad peaks are represented by squares, clear peaks by triangles, double peak shapes by twin triangles and plateau shapes by half squares.

It is remarkable how all HVSRs have their peaks in the frequency band below or close to 1 Hz, independent of its shape. Such low-frequency resonances give information of the deep interfaces between Quaternary-Neogene sediments and soft rocks and Upper Miocene-Triassic stiff rocks. The HVSR curves along sections a-a' and b-b' are shown in Figure 10. The overall trend of frequencies to decrease towards the SE (Figs. 9, 10b) is consistent with the increments in depth shown in the geological cross sections (Fig. 2b) as well as in isopach maps of the basin built from borehole analysis [32].

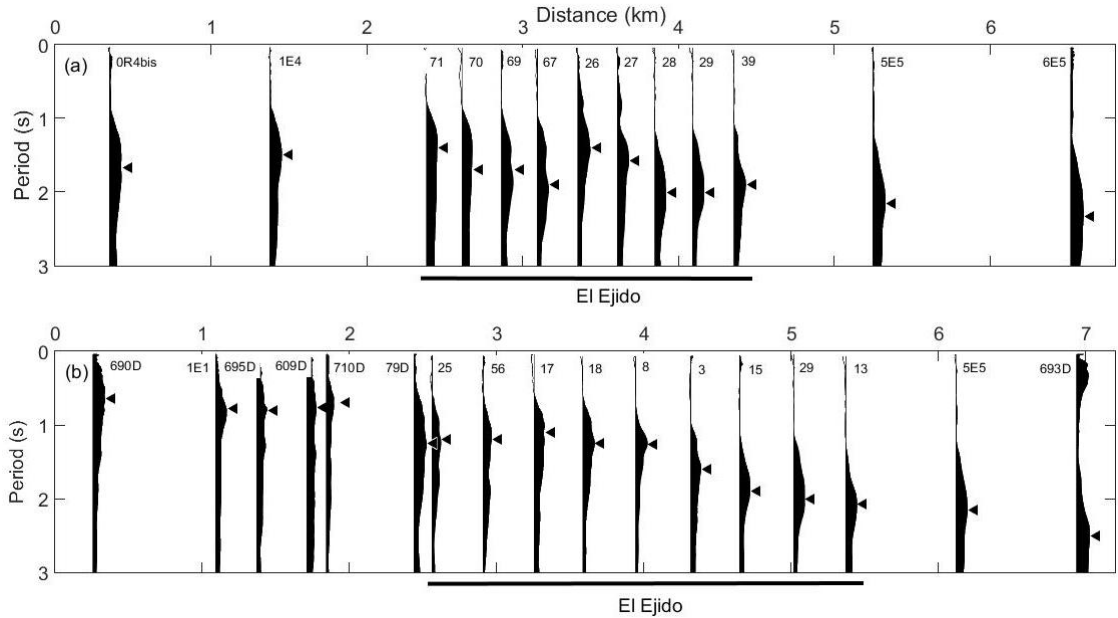

**Fig. 10.** Representation of HVSRs along sections a-a' (panel a) and b-b' (panel b). These profiles have been extended beyond the town limits in order to oversee consistency with the general basin structure. Peak amplitudes have been normalized and their corresponding fundamental periods are highlighted with black triangles.

### 3.3.1. Inversion of HVSR shape

An innovative interpretation of the HVSR inspired by the possibility of retrieving the 3D elastodynamic Green's tensor between two stations embedded in an elastic medium from cross-correlation of their ambient noise records (ambient noise interferometry) was introduced by Sánchez-Sesma et al. [29]. The theoretical basis of this general hypothesis was developed in several research works [53–55] and confirmed in experiments with ambient noise measurements [56].

In the underlying theoretical approach, a 3D-diffuse equipartitioned displacement vector field is assumed to be established within an elastic medium with a bounded heterogeneous region. Then, power spectral densities of motion along any Cartesian axis $m$ at an arbitrary point **x**, $P_m(\mathbf{x}; \omega)$, are proportional to the imaginary part of the corresponding Green's function at the source $G_{mm}(\mathbf{x}; \mathbf{x}; \omega)$, where $\omega = 2\pi f$ is the circular frequency and $m = 3$ stands for vertical direction [57]. Taking the usual definition, the H/V spectral amplitude ratio under DFA is given by

$$HVSR(x; \omega) \equiv \sqrt{\frac{2P_1(\mathbf{x};\omega)}{P_3(\mathbf{x};\omega)}} = \sqrt{\frac{2Im[G_{11}(\mathbf{x};\mathbf{x};\omega)]}{Im[G_{33}(\mathbf{x};\mathbf{x};\omega)]}} \;. \tag{1}$$

According to this theory, the inversion of seismic velocity models from the experimental HVSRs has been carried out considering a stratified elastic halfspace by using algorithms described in [30] which take advantage of compact expressions of the imaginary part of Green's functions [58] for the special case of coinciding source and receiver at the top of the layered media. "HV-Inv" software (https://w3.ual.es/GruposInv/hv-inv/, [30,59]) was used to obtain seismic velocity profiles for sections a-a' and b-b' (Figure 4). The former profile was chosen according to clarity of the intersected peaks as well for being in the flatland, where shallow exploration methods found softer materials, while the latter by criteria based on the expected trend of sediment thickness growing. In general, the HVSRs were considered between 0.25 Hz and 15 Hz. This frequency range was shortened in cases of unstable behaviour at low frequencies (see e.g. Fig. 8c below 0.4 Hz) and in some cases presenting singular features near this high-frequency limit (e.g. flat HVSR with amplitudes below 1), probably related with very shallow and local ground characteristics. This latter simplification avoids excessive and site-depending model parametrization in the upper 10-15 metres.

Attending to the stratigraphic models, where five main geological units are distinguished (Fig. 2b), the inversion procedure began with the establishment of parameters ranges for a general model of four layers over a half-space (Paleozoic-Triassic unit). The limits for the seismic velocities, density, thickness and Poisson's ratio of each layer (Table 4) have been defined on the basis of previous geophysical information (e.g. [33,37]) besides a calibration through 20 HVSR measurements on boreholes, where stratigraphy is known. A combined use of Monte Carlo sampling (in a first stage) and downhill simplex optimization was the strategy followed to adjust the theoretical HVSR curves, resulting from trial models in the parameters space (Table 4), to the experimental ones. Likelihood of the tested models, for both the HVSRs and the dispersion curve inversions (section 3.4) is based on the definition:

$$misfit(\boldsymbol{m}) = \frac{1}{n}\sum_{i=1}^{n}\frac{\left(X_{obs}(\omega_i) - X(\omega_i, \boldsymbol{m})\right)^2}{\sigma_X^2(\omega_i)}, \qquad (2)$$

where $X$ is the inverted observable, $\sigma_X$ is its standard deviation, and subscripts *obs* and *th* stand for observed and theoretical values.

An example of inversion at a borehole site, using depth constraints based on stratigraphy, is shown in Fig. 11. Conversely, results for sites S1, S2 and S3 (Fig. 4) are shown in Fig. 12 as examples of inversions of HVSR where *a priori* information is lacking, using the general parameter ranges presented in Table 4. These three curves show significant peaks with fundamental frequencies below 0.7 Hz and amplitudes between 4 and 5, reproduced by models with clear velocity contrasts between 400 and 650 m deep. These contrasts can be interpreted as interfaces with hard calcareous rocks. The HVSRs become relatively flat above the main peak except for site S1, where a wide bump in the band 1.2 – 10 Hz evidences a more complex shallow structure.

**Table 4.** Ranges for model space exploration used in the inversion procedure of HVSR curves

| Layer | Thickness (m) min – max | $V_P$ (m/s) min – max | $V_S$ (m/s) min – max | $\rho$ (kg/m³) min – max | $\nu$ min – max |
|---|---|---|---|---|---|
| 1st | 1 – 70 | 500 – 1900 | 330 – 900 | 1950 - 2050 | 0.15 - 0.35 |
| 2nd | 20 – 300 | 800 – 2100 | 500 – 1000 | 2000 – 2200 | 0.21 - 0.35 |
| 3rd | 20 – 300 | 1400 – 2500 | 850 – 1200 | 2200 – 2400 | 0.21 - 0.35 |
| 4th | 30 – 300 | 1650 – 4000 | 1000 – 1900 | 2200 – 2600 | 0.21 - 0.35 |
| Half-space | | 2450 – 5100 | 1500 – 2800 | 2700 - 2750 | 0.21 - 0.28 |

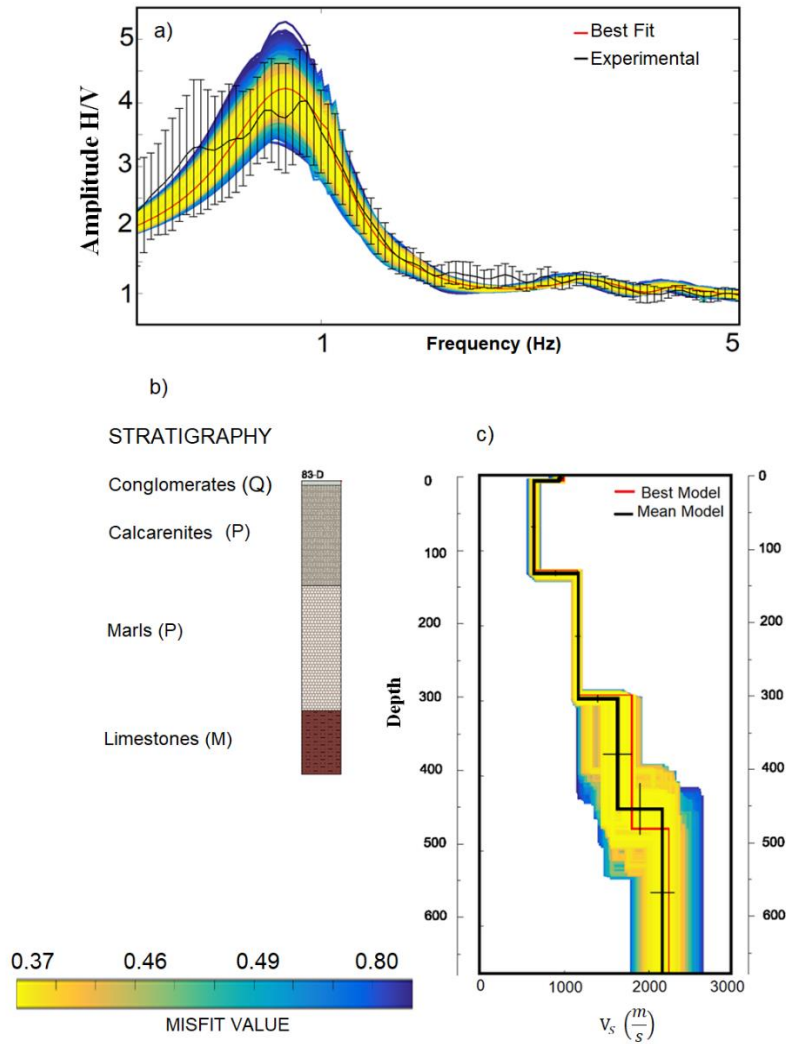

**Fig. 11.** Example of HVSR inversion on borehole 83-D (see location in Fig. 2a). a) Experimental HVSR curve; b) Stratigraphic column; c) Obtained $V_S$ profile.

The 1-D seismic velocities models obtained along lines a-a' and b-b' were assembled into pseudo-2D profiles (Fig. 13) by using a radial-basis function interpolation through "Surfer" software. The a-a' cross-section (Fig. 4) arises from the best fits of 7 HVSR curves distributed along 1800 m, and the second one (b-b' Fig. 4) from 9 curves along 2800 m.

In the light of these two profiles, various remarks can be made about the sediments drift and seismic velocity ranges attributed to each geological unit. Firstly, it is observed how the depth down to the halfspace increases from 370 m in the northwest limit to 620 m in the southeast end (b, b' respectively) while it keeps around 550 m deep along a-a' profile. Since overlapping between velocity values of different materials is accepted, an unequivocal and direct identification of each layer from these profiles is not possible. Thus, we prefer to give results in terms of averaged values for each layer of the general model established for inversion (Table 5).

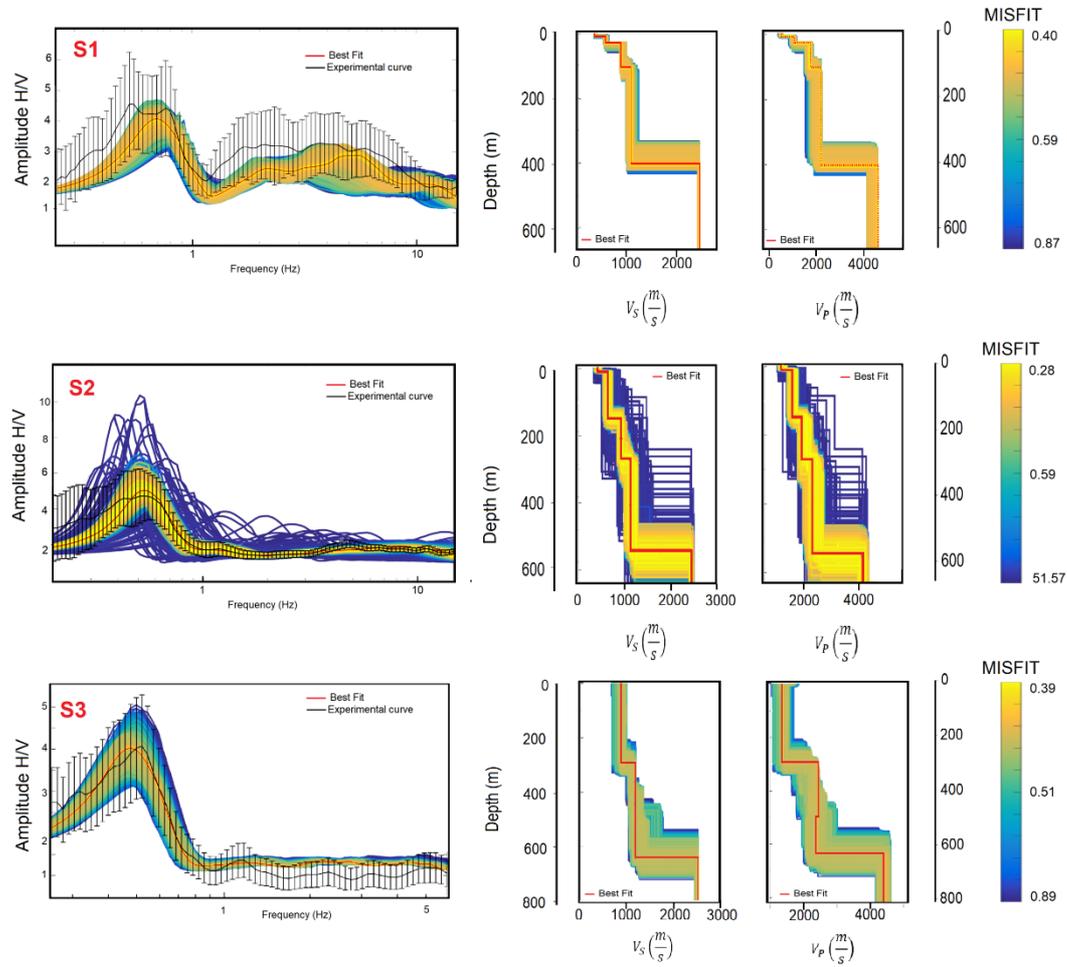

**Fig. 12.** Inversion of HVSRs (black lines) on sites S1, S2 and S3 (see locations in Figs. 4 and 14). Theoretical curves and their corresponding tested models are coloured according to the misfit bars. The best model achieved and its theoretical HVSR are highlighted in red.

**Table 5.** Mean values and standard deviations obtained from the 16 best fits used for constructing the 2-D profiles shown in Fig.13

| Layer | Thickness (m) | σ (m) | $V_S$ (m/s) | σ (m/s) | $V_P$ (m/s) | σ (m/s) |
|---|---|---|---|---|---|---|
| 1st | 14 | 13 | 483 | 70 | 851 | 111 |
| 2nd | 110 | 57 | 688 | 80 | 1251 | 164 |
| 3rd | 144 | 47 | 990 | 100 | 1936 | 222 |
| 4th | 234 | 112 | 1381 | 184 | 2694 | 486 |
| Half-space | - | - | 2255 | 329 | 3950 | 602 |

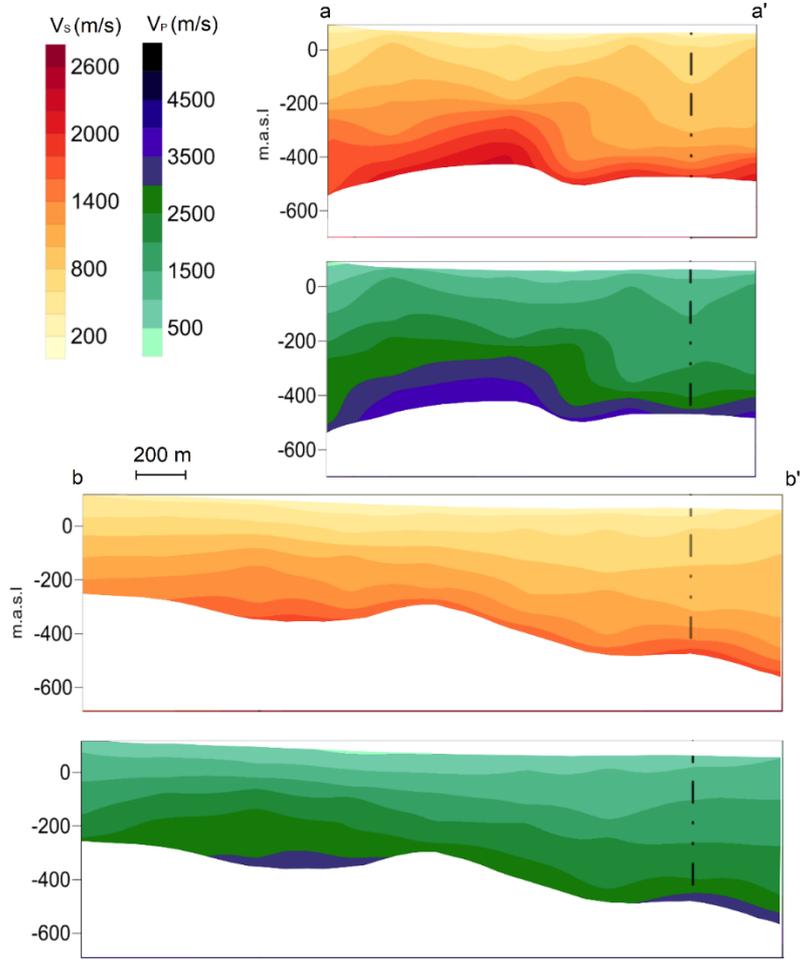

**Fig. 13.** Interpolated 2-D cross-sections showing the deep structure for El Ejido town in terms of seismic velocities, built from interpolation of 1-D models resulting from HVSR inversion. Models are shown down to the inverted halfspace depth. The vertical black line represents the intersection of profiles a-a' and b-b'.

*3.4. SPAC technique*

The second procedure used here to define the shallow and deep shear-wave velocity structure in El Ejido was based on analysis of dispersion curves of fundamental-mode Rayleigh waves retrieved from the SPAC method [22]. The SPAC coefficient $\rho_{SPAC}(\omega, R)$ was computed as the azimuthally-averaged cross-spectral densities between the records of seismometers evenly spaced on a circumference and a central station, using the power spectral density at the centre for normalization. The SPAC coefficient was separately computed for a set of overlapping time windows, plotted as a function of time and frequency and averaged discarding windows with evident disturbances. Finally, the phase-velocity $c_R(\omega)$ was computed for each frequency from equation (3)

$$\rho_{SPAC}(\omega, R) = J_0\left(\frac{\omega}{c_R(\omega)} R\right), \qquad (3)$$

where $J_0$ represents the zero-order Bessel function.

The next stage is the fitting process of $c_R(\omega)$ by means of an inversion procedure to obtain a 1D layered model profile, in terms of body waves velocities and densities, whose theoretical dispersion curve approximates closely the experimental one.

SPAC observations were carried out in three open spaces inside the urban area (Fig. 14). Pentagonal or triangular arrays with several apertures were deployed at each site. The radii ranged from 5 to 40 m for sites S1 and S3 and from 120 to 310 m for S2 (Table 6). Broadband sensors were used as independent stations to record ambient noise in the largest array (site S2) meanwhile VSE-15D sensors all connected to a single SPC-35 digitizer were used for smaller radii (sites S1 and S3). The former setup has the advantage of wireless recording over the latter, being used here in larger arrays. The VSE-15D velocimeters have a flat frequency response from 0.25 to 70 Hz. The records were sampled with a rate of 100 sps, lasting half an hour for the small apertures (sites S1 and S3) and one hour for the larger ones (S2). The window lengths were defined to ensure that they contained each analysed period 50 times at least. Rayleigh wave velocities were retrieved in the range from 9 to 28 Hz for small arrays (S1, S3) and from 0.7 to 7 Hz for the big one (Fig. 6).

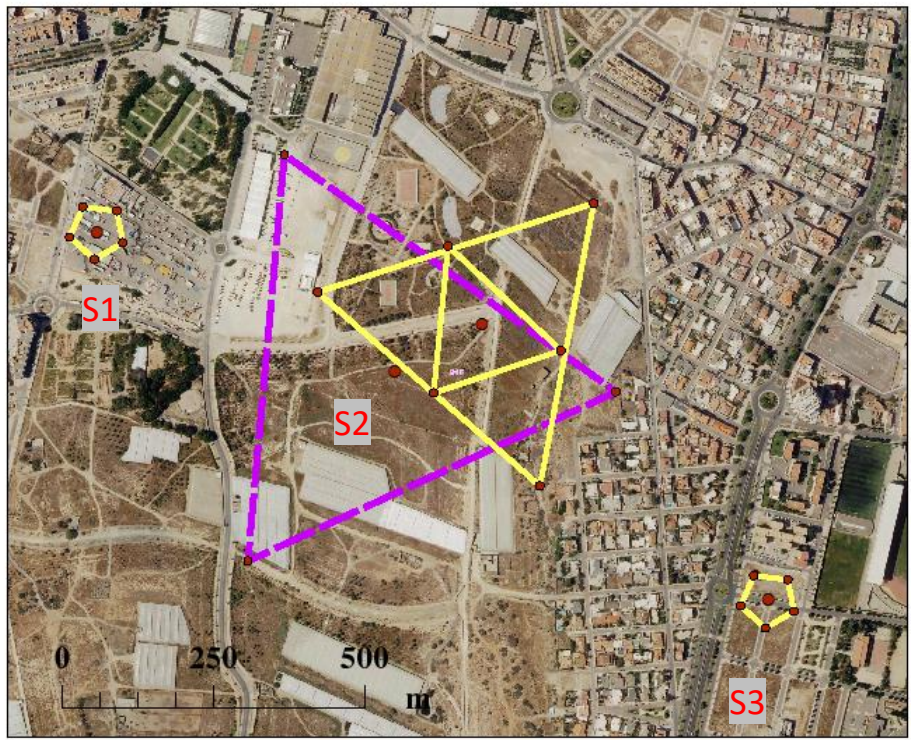

**Fig. 14.** Array setup for SPAC observations in El Ejido. Red points are the sensor positions on vertexes and centre of each array geometry deployed. In the case of pentagonal arrays, only the largest radius is shown.

**Table 6.** Summary of array features and ranges of dispersion curves retrieved with the SPAC technique

| Site | Radii (m) | $\Delta f$ (Hz) | $\Delta c_R$ (m/s) | $\Delta V_S$ (m/s) | $\Delta z$ (m) | $V_{S30}$ (m/s) |
|---|---|---|---|---|---|---|
| S1 | 5 – 10 – 20 – 40 | 9 – 24 | 402 - 816 | 235 - 1066 | 3-32 | 578 |
| S2 | 120 – 220 – 310 | 0.7 – 7 | 514 -1842 | 429- 2195 | 12-950 | 514 |
| S3 | 6 – 12 – 24 | 12 – 28 | 523 - 788 | 391-1045 | 2-27 | 705 |

Using the assumption of an effective sampling depth between $\lambda/3$ and $\lambda/2$ for fundamental-mode Rayleigh waves of wavelength $\lambda$ [39,60], the maximum resolved depth would be between a third and a half of maximum wavelength ($\lambda_{max}$) which can be estimated from the velocity associated to the minimal frequency $f_{min}$ of the dispersion curve as $\lambda_{max} = c_R(f_{min})/f_{min}$. Thus, the maximum estimated depths are between 22 and 33 m for site S3, and from 30 to 45 m for site S1. The big tringle at site S2 would reach maximum depths between 880 and 1320 m.

The results of the inversion procedure are displayed in Fig. 15. Inversion-derived values of $V_{S30}$ can be calculated from thicknesses and $V_S$ values of the best-fit models (Table 6). Due to the limited resolution of the large array (site S2) at short wavelengths, the definition of the upper layers could be fuzzier, even though the computed $V_{S30}$ value remains in a reasonable range in comparison with the smaller arrays. These models are consistent with the results of the MASW, reaffirming the moderate spatial velocity variations in the scale of hundred metres and increasing the exploration depth reached with that active technique (slightly for S1 and S3 and substantially for S2). However, the inversion of Rayleigh wave dispersion curves is sensitive to an overall or smoothed trend of $V_S(z)$, implying a degree of uncertainty in some characteristics of the velocity structure (e.g. existence of sharp velocity contrasts). The ground model can be more constrained by adding other observations such as the HVSR, which are very sensitive to the existence and depth of such interfaces.

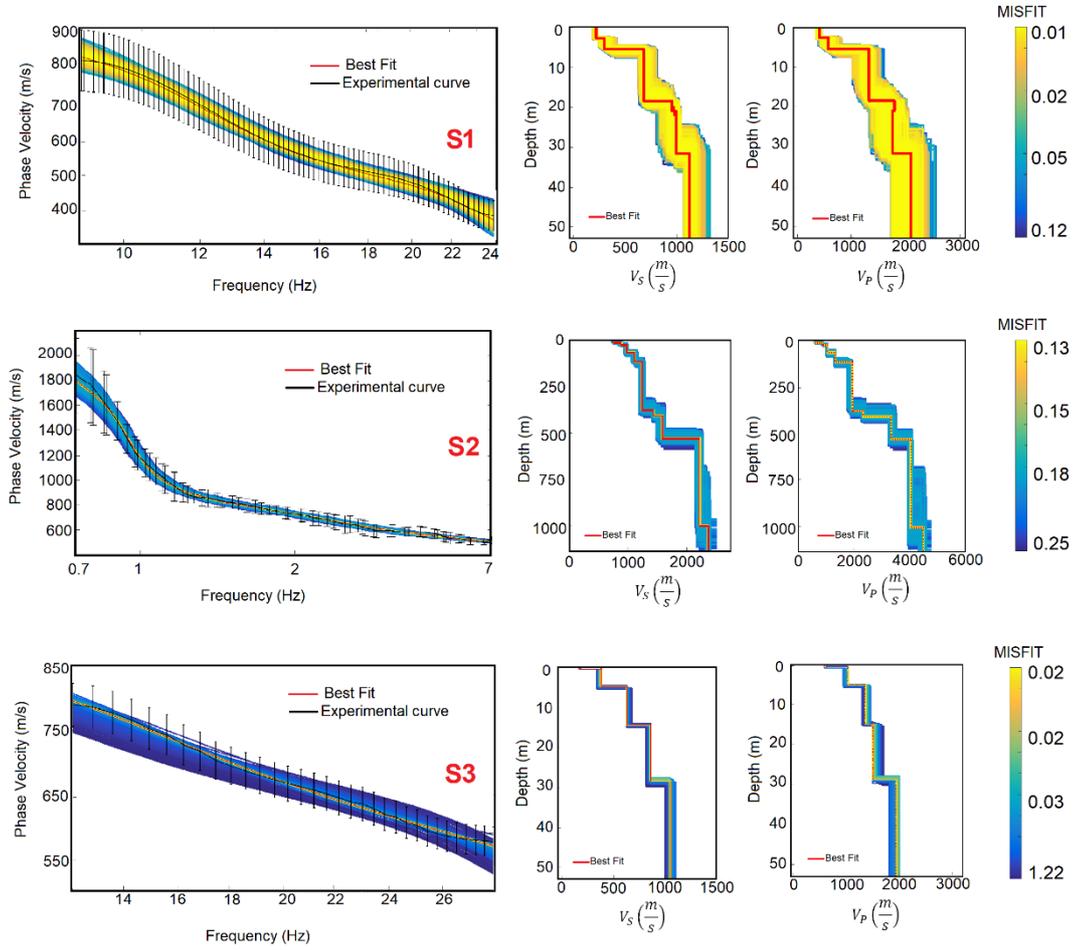

**Fig. 15.** Empirical dispersion curves (black lines) and theoretical curves corresponding to the tested models (coloured lines). Body wave velocity models for each analysed site. Best model achieved is highlighted in red.

**3.5. Joint Inversion of HVSR and dispersion curves**

Dispersion curves from SPAC and HVSR at the array centres were processed together by a joint inversion scheme (Fig. 16). For site S2, the dispersion curve was extended to high frequencies (48 Hz) including the mean dispersion curve we got from MASW profile P 3.3 (Fig. 6), which go across the largest triangle (Figs. 4 and 14). "HV-Inv" software was also used in this final stage in order to characterize the S-wave velocity structure. Suitable weights for the contributions of HVSRs and dispersion curves were used, defining the misfit of a model **m** as

$$misfit(\mathbf{m}) = \frac{1}{2n}\sum_{i=1}^{n}\frac{(HVSR_{obs}(\omega_i) - HVSR_{th}(\omega_i, \mathbf{m}))^2}{\sigma_{HVSR}^2(\omega_i)} + \frac{1}{2m}\sum_{i=1}^{m}\frac{(c_{obs}(\omega_i) - c_{th}(\omega_i, \mathbf{m}))^2}{\sigma_c^2(\omega_i)}, \quad (4)$$

where *n* and *m* are the number of samples for the HVSR and the dispersion curve, respectively. This choice equalizes the weight of both observables in the misfit, regardless the respective number of samples (frequencies) and it does not appear to slow down the convergence of the algorithm.

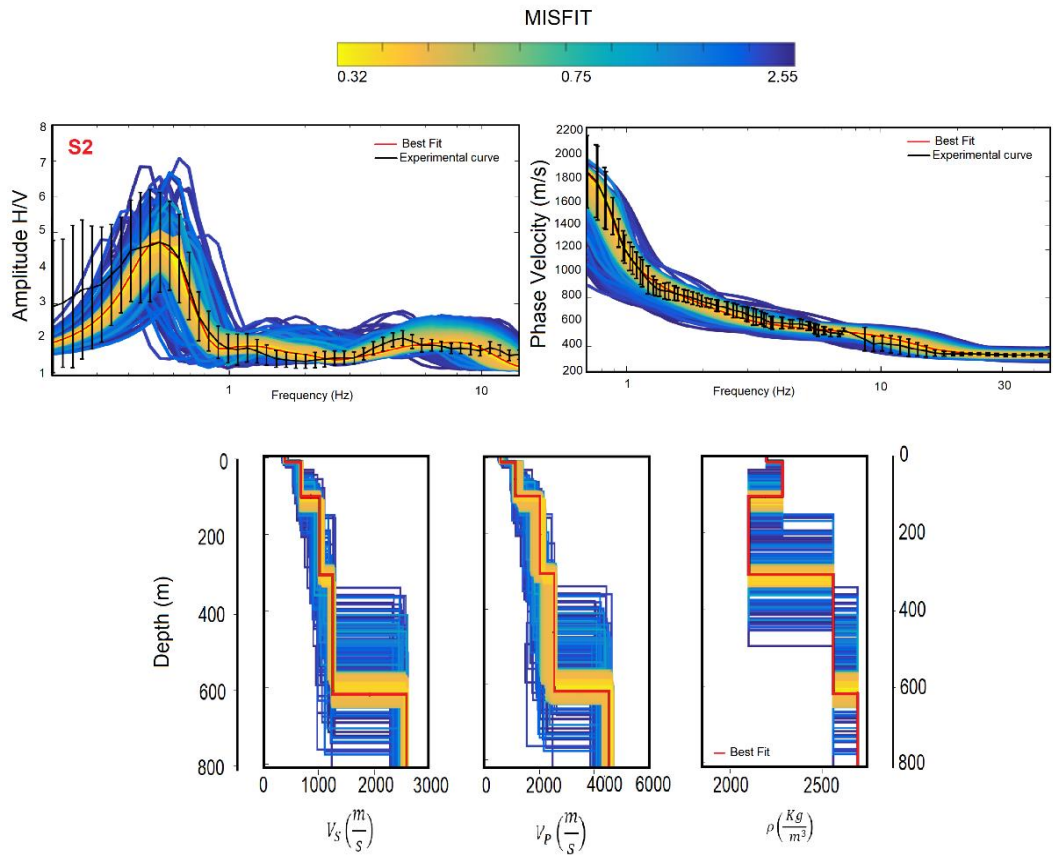

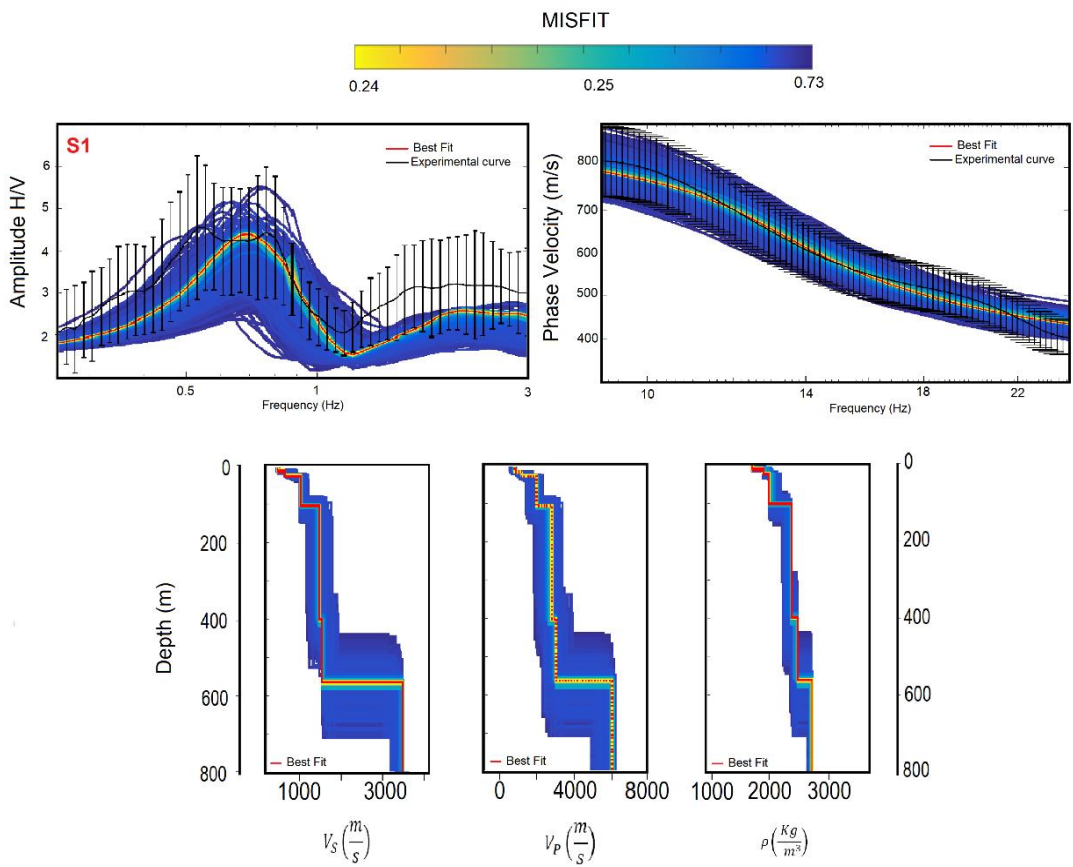

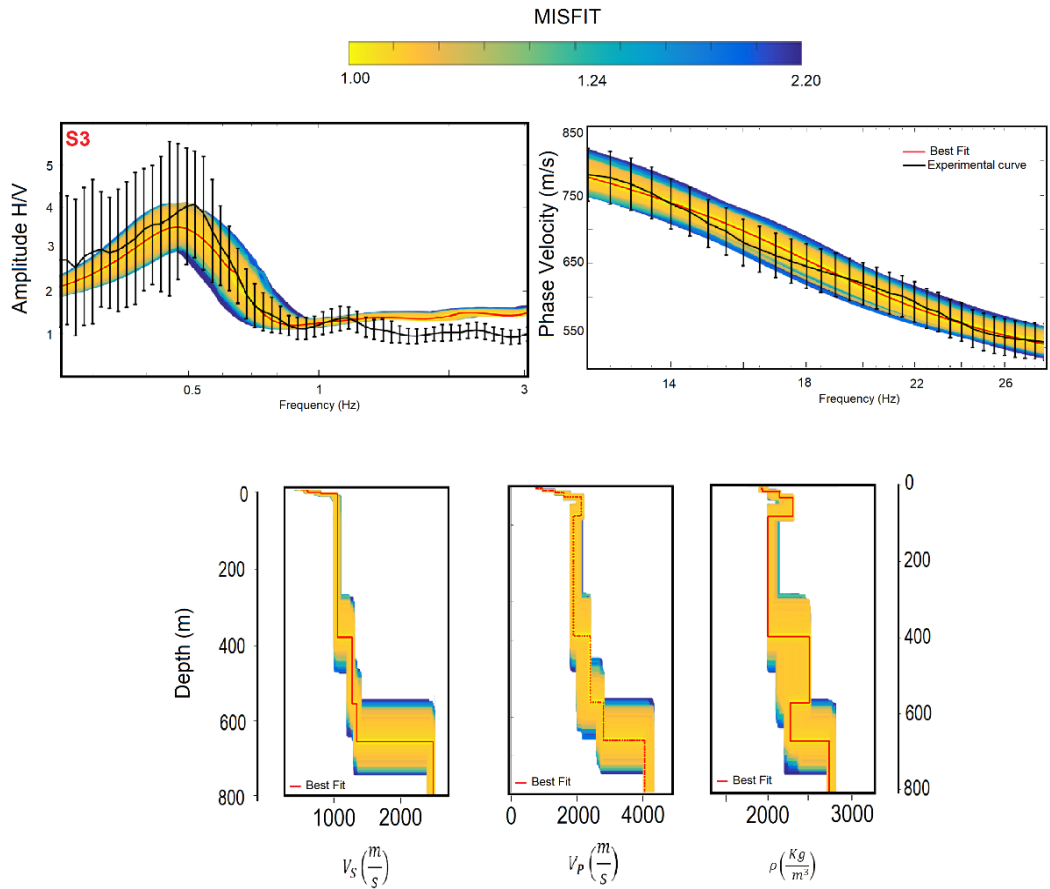

**Fig. 16.** Results of joint inversions from empirical HVSR and dispersion curves (black lines) for each analysed site (Fig. 14). The theoretical curves are coloured as their corresponding velocity and density models. Best model achieved is highlighted in red.

The minimum frequency of the dispersion curve retrieved at site S2 approaches the fundamental frequency given by the HVSR technique, providing additional information about the deep sedimentary structure. On the other hand, the inclusion of the dispersion curve derived from MASW, which provided a theoretical exploration depth of about 23 m, improved the model accuracy in the upper two layers. The major peak in the theoretical HVSR is mainly controlled by the deepest velocity contrast of these inverted models. $V_S$ models obtained by joint inversion of HVSRs and Rayleigh wave dispersion curves have a slightly stiffer half-space in comparison with inversion of HVSR only (Fig. 12). Results for S2 show a somewhat deeper and stiffer basement in comparison with the model derived from dispersion curve fitting (Fig. 15). In general terms, joint inversions provide more constrained models, therefore these halfspace properties should be considered to have been determined in a more reliable procedure. Parametric analysis gives evidences on how cost functions (misfit) display sharper minima for joint inversion [59]. An example of this is shown in Figs. 17a-c, where the misfits for HVSR, $c_R$ (Eq. 2) and joint inversion (Eq. 4) are compared for some perturbations in the model for S2 inverted in Fig 16, which is represented by a black dot. Along the horizontal axis, velocities and thickness of the upper four layers of the model are rescaled by the same factor, preserving the travel time of P and S waves through the structure. $c_R(f)$ shows high sensitivity to these perturbations which involve variation of velocities in shallow and intermediate layers. The vertical axis corresponds to variations in P and S velocities for halfspace, preserving the Poisson's ratio. In this case, the HVSR presents greater variations. Joint inversion (Fig. 17c) provided real sensitivity to velocity values and constrained better the existing deep impedance contrasts. It should be noted that the velocity contrast between bedrock and sedimentary fillings is a critical parameter for assessment of strong-motion amplifications. In particular, ground amplifications of 3 times at the fundamental frequency can be estimated at site S2 for vertically incident S-waves (it corresponds to 6 times normalizing by the amplitude of the incident wave, as in Fig. 17e).

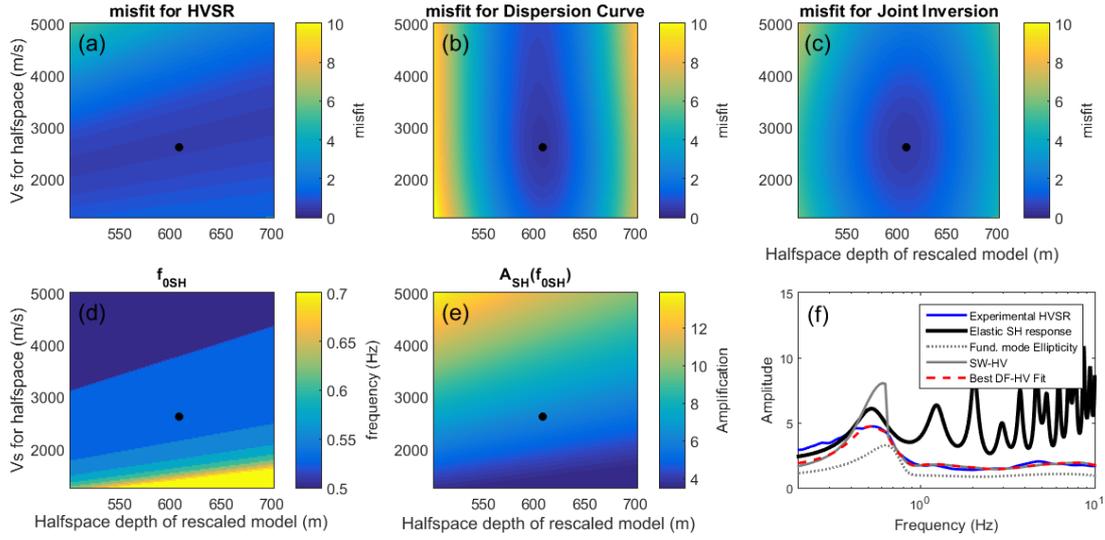

**Fig. 17.** (a-c) Misfit values considering HVSR, $c_R$ (Eq. 2) and joint inversion (Eq. 4), respectively, for perturbations in the model inverted for site S2 in Section 3.5 (Fig 16). Along the horizontal axis, velocities and thickness of the upper four layers of the model are rescaled by the same factor. Vertical axis corresponds to variations in halfspace velocities, preserving the Poisson's ratio. Black dots represent unperturbed parameters. (d-e) Evolution, under the same model perturbations, of the fundamental frequency for vertically incident S waves ($f_{0SH}$) and the amplification at $f_{0SH}$ with respect to the amplitude of the incident wave. (f) Experimental HVSR at S2 compared with the transfer function of vertically incident S-waves computed for the inverted model, with the fundamental-mode Rayleigh wave ellipticity and with the theoretical HVSR under the DFA for full-wavefield and for surface waves only.

The HVSR(f) function depends only on the travel time of P and S waves within the layers and on density contrasts, so that *equivalent* models could be generated if all thicknesses and velocities (or all densities) were rescaled by the same factor. For instance, Sánchez-Sesma et al. [29] interpreted a HVSR curve from Texcoco (Mexico City), similar in shape and predominant frequency (0.47 Hz) to some found in El Ejido, as the resonance of 40 m of very soft clays ($V_s$ as low as 70 m/s) overlying a much stiffer substratum (1000 m/s). Thus, constraints to the shallow layer velocities provided by high-frequency segments of the Rayleigh wave dispersion curves, as those retrieved in S1 and S3, still contribute to the non-uniqueness reduction of the resulting models.

## 4. Discussion and Conclusions

Five main types of materials have been identified in El Ejido town from analysis of deep boreholes. Lithologies range from Quaternary deposits on surface, which are mainly composed by sands, silts, gravels and clays to Paleozoic-Triassic dolomites (Fig. 2b). Results of geotechnical tests provided lithological information down to 50 metres deep. The Quaternary materials show a relatively high average $N_{SPT}$ value of 32 blow counts. Interleaved in the first metres of Quaternary deposits, calcareous crusts and conglomerates are found, causing refusal in SPT logs. Density of Quaternary materials varies between 1.72 g/cm$^3$ for gravels and 1.94 g/cm$^3$ for clays. Pliocene calcarenites show an average $N_{SPT}$ value of 49 blows and a density of 2.15 g/cm$^3$. Materials such as gravels, conglomerates, calcareous crusts and calcarenites often show similar $N_{SPT}$ values. On the other hand, some materials (mainly clays and silts) presented $N_{SPT}$ varying in wide ranges, so that lithology provides only a rough indication of the physical properties of soils. Empirical relationships (Tables 1 and 2) estimate $V_S$ values for Quaternary materials varying in a range from 205 to 379 m/s, while calcarenites, which only outcrop on southern neighbourhoods (Fig. 2a), would have $V_S$ of 511 ± 150 m/s.

More detailed $V_S$ profiles for the shallow layers have been obtained by using the MASW method. Figure 18 shows examples of these profiles with superimposed geotechnical information. In terms of Eurocode 8, the ground in El Ejido urban area can be classified as type B, whose $V_{S30}$ values are established to range between 360 and 800 m/s. MASW cross-sections reveal that materials with $V_S$ within this range are predominantly found in the upper 10 – 15 m. Some layers and bodies of softer materials ($V_S$ < 360 m/s) with different sizes and thicknesses appear interleaved or atop these units in the central-southern area (West of MASW profiles P2, line P3 and East of P4) and on the Quaternary alluvial cones of the western zone (P6 line, Fig. 7). The central-southern area presents the softest profiles, being also the only zone of the town where the HVSR show peaks of moderate amplitude (2 – 3) at intermediate frequencies (6 - 8 Hz), numerically coinciding with theoretical S-wave resonances in the upper 10 - 20 m. Velocities above 600 m/s are mainly reached within layers of Pliocene calcarenites (e.g. SPT 8, Fig. 18b) or in levels described as marly or stiff clays (e.g. SPT loggings 9, 10, 12, Figs. 18a, b). As shown in Fig. 18, a good agreement between alternative methods has been reached where comparison was possible.

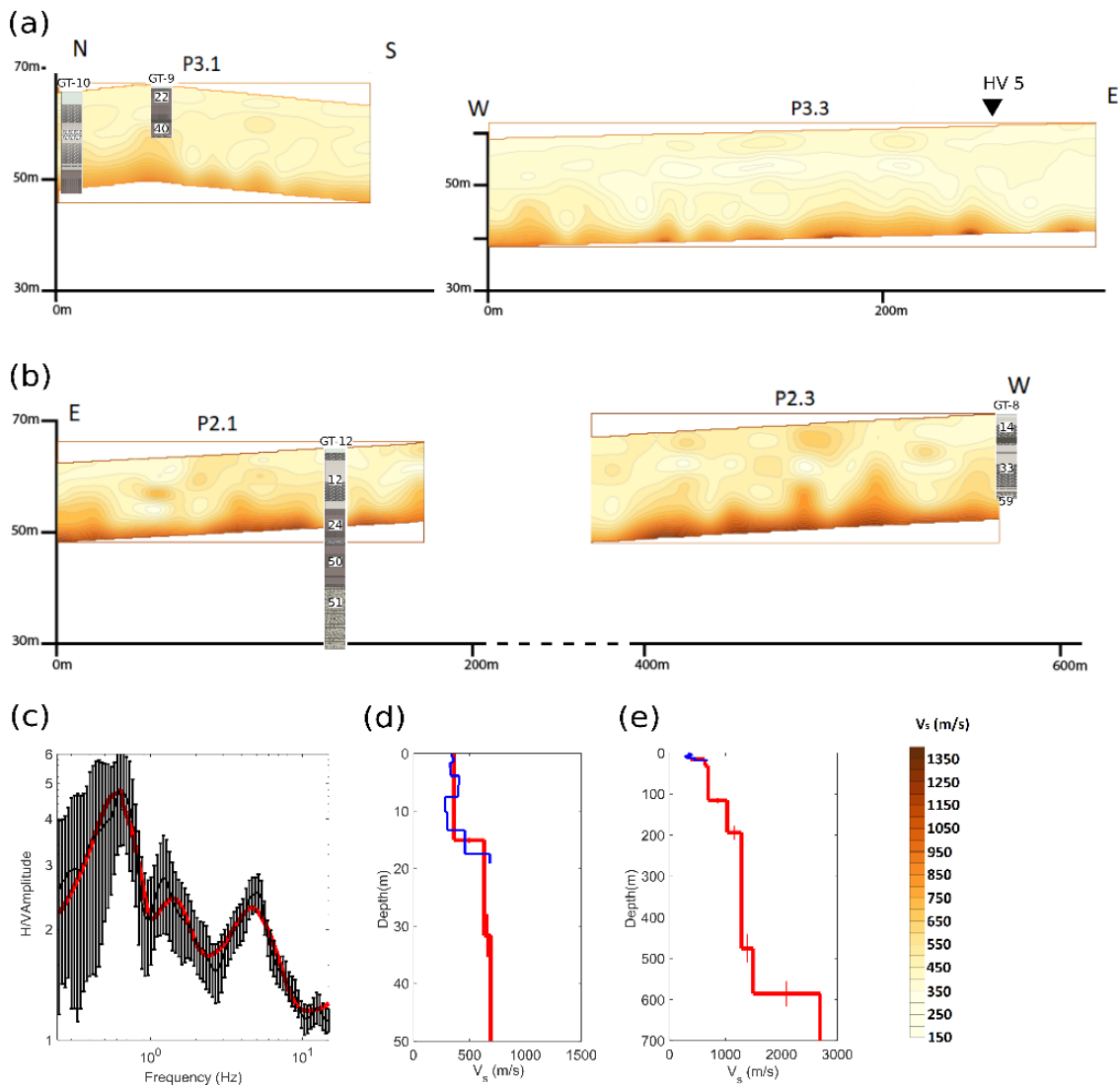

**Fig. 18.** Examples of $V_S$ cross-sections with superimposed stratigraphic columns obtained at nearby geotechnical profiles. (a) Part of profile P3 (b) Part of profile P2. Average $N_{SPT}$ values of the strata are shown. (c) Experimental (black line) and modelled (red line) HVSRs for site HV5, exhibiting a secondary peak at 5 Hz. (d) Upper layers of the $V_S(z)$ model inverted from HVSR at HV5 (red line), which belongs to section a-a' (Figs. 4 and 13), and shallow profile obtained from MASW at the same point (blue line). Models are shown down to the basement in panel (e).

Seismic methodologies present certain complexity and costs related with the acquisition and deployment of the equipment. It consisted of 50 m long linear array for MASW measurements and, in the case of SPAC measurements, of several bidimensional arrays with different apertures that need to be placed during several hours in a quiet area. With this in mind, the MASW dataset acquired in El Ejido can be used to check and to adjust for this zone a simplified seismic method for estimation of local $V_{S30}$ values using simpler arrays, suitable for active and passive sources. This approach could be useful in the future to characterize this town with higher sampling density or to extend this study to surrounding areas. Konno and Kataoka [61] suggested that the experimental setup used for determination of $V_{S30}$ can be simplified if the array is designed to operate in suitable narrow wavelength ranges. They proved empirically the practical equivalence between $V_{S30}$ and the Rayleigh wave velocity for wavelengths between 35 and 40 m ($c_{35}$, $c_{40}$) for a large dataset in Tokyo area.

In the present work, during the checking procedure of the advantages related to this faster approach it has been observed how $V_{S30}$ is better estimated by $c_{45}$ or $c_{50}$ velocities (Figure 19). This result agrees with those pointed out for places with deep water tables (e.g. [42]) and it can be partially explained by the difference between the characteristic Poison's ratio of soils in these two places (about 0.25 for El Ejido, corresponding to dry and relatively hard materials, and above 0.45 for Tokyo). The much smaller Poisson's ratios reported in our explored area would lead to higher inverted $V_{S30}$ values for a given dispersion curve (up to ~4% increment), as a result, the equivalence with the increasing function $c_R(\lambda)$ would be reached for longer wavelengths.

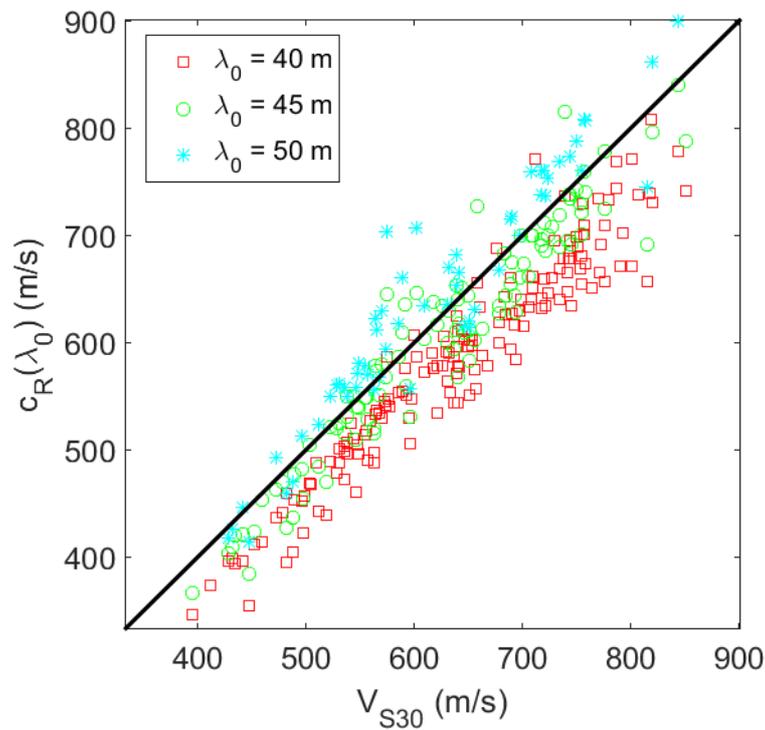

**Fig. 19.** Phase velocity of Rayleigh waves for fixed wavelengths $\lambda_0$ vs. $V_{S30}$ for the individual 1D models which constitute the MASW profiles. All dispersion curves reaching the target wavelength and the corresponding 1D $V_S$ models are considered.

The HVSR method has been applied in order to obtain a map of predominant frequencies and shape typologies from 94 sites in the town distributed on a 250 x 250 m grid. Robustness of this technique to provide the fundamental frequency when clear HVSR peaks are observed is widely accepted in the literature. That case corresponds to strong contrasts of seismic velocities between soft sediments and bedrock [47,48]. The obtained results show the effectiveness of the HVSR method even in conditions of relatively high-velocity Pliocene and Miocene rocks keeping an important contrast with the underlying materials. Clear peaks in El Ejido are mainly concentrated in the southeast part, with peak frequencies between 0.5 and 0.7 Hz, although some are scattered between broad- and double-peak neighbours with frequencies from 0.7 to 1 Hz. Broad peaks are the highest percentage of the HVSR dataset, with lower cut-off limits between 0.3 and 0.7 Hz and higher limits from 0.6 to 1.2 Hz. This latter value is specially concentrated in the northern row of the grid. Double-peaked HVSRs are sparsely distributed in the town. In this case, the difference in frequency between the peaks is often moderate so that they appear partially overlapped, with the low frequency subpeak between 0.4 and 0.6 Hz and high frequency one between 0.6 and 0.9 Hz. Plateau-shaped HVSRs [49] are also located in northern rows, with cut-off frequencies ranging from 0.8 to 1.2 Hz. Predominant frequency distribution is not well-linked to shallow soil conditions. On the contrary, a clear dependence on the total thickness of sedimentary materials, estimated from deep borehole and seismic reflection data, is found in the study area. The highest predominant frequencies were found in the NW area of the town, whereas the SE zone shows clear HVSR peaks with lower frequencies. In case of double or broad peaks, the high-frequency value accords better with the nearby clear peaks and with the overall NW-SE trend of the periods to increase.

Since HVSR curves are sensitive to the sediment-bedrock interface, they have been used as a passive technique for approximate subsoil mapping [62–65]. In this work, the inversion of HVSR curves under the DFA provided us with 1-D velocity models where the sediments-bedrock interface is mapped (Fig. 12) as well as other shallow secondary contrasts (e.g. Fig. 18d, e). In this case, the heterogeneity in the upper layers revealed by the geotechnical tests is highly simplified in terms of an equivalent model. In fact, the upper 20 m were described by a maximum of 2 layers out of the 5 defined for the inversion process (Table 4). Non-uniqueness of HVSR inversions was diminished providing additional constraints to the model parameters. In particular, an essential part of the procedure was a previous calibration through measurements performed next to deep boreholes, in which the interfaces described in previous geological studies of Campo de Dalías were identified. This processing method based on the DFA is a simple way to achieve reliable 1-D profiles from the fitting of experimental HVSR curves without the need to extract a particular type of waves (see e.g. [66] for an alternative method based on the selection of Rayleigh wave arrivals). Since effects of Love waves, higher modes of Rayleigh waves and body waves in the HVSR are not always negligible (e.g. [67,68]), full-wavefield inversions are definitely more suitable than oversimplified approached based on modelling the HVSR of noise by the Rayleigh wave ellipticity (e.g. [26, 69]). For example, the peak frequency for the inverted model S2 shifts from 0.52 Hz to 0.63 Hz if only fundamental-mode Rayleigh waves are considered in the forward modelling (Fig. 17f). Moreover, body waves can be the main contribution to the vertical motion in some frequency bands with low power or high ellipticity of the Rayleigh waves, preserving realistic HVSR amplitudes in the forward modelling (e.g. compare red and solid-grey lines in Fig. 17f).

The assembling of individual velocity profiles (Fig. 13) has confirmed the south-eastward increase of sediment thickness, overcoming the poor information on the deep structure within the urban area. Active-source reflection surveys carried out in Campo de Dalías show that El Ejido lies in the north flank of an asymmetric syncline with ENE-WSW trend. N-S reflection profiles showed a mean apparent dip of about 8º for that flank. Similar behaviour is observed in the two profiles constructed for the urban area by inversion of HVSR curves. The NW-SE section (profile b-b' Figs. 4, 13) shows an apparent dip of around 7º with a maximum depth of 620 m down to the half-space, which is close to the 603 m value attributed by P-wave analysis of an active line that passes near b'. Inversion of broad and double peaks in the HVSR often led to models with significant velocity contrasts at intermediate depths and moderate contrasts between the halfspace and the overlaying materials (e.g. points 25 and 56 in the NW side of profile b-b', Figs. 10 and 13). On the other hand, the E-W section in El Ejido town (profile a-a' in Figs. 4, 13) shows variations of the half-space depth between 490 and 630 m. Such geometry also agrees well with a parallel reflection profile conducted 1.8 km northwards (see AL-04 and AL-01 profiles in [70]).

Joint inversions of Rayleigh-wave dispersion curves and HVSR curves were carried out with the aim of enhancing the reliability of the velocity models, reducing the non-uniqueness associated to the independent inversion of any of these curves. An array of larger aperture (site S2, Figs. 4 and 14) was deployed to retrieve Rayleigh wave velocities at low frequencies, close to the HVSR predominant frequency. Even though the resulting ground model has a more constrained deep structure, mainly in $V_S$, the obtained seismic velocities are still within the ranges defined for the models inverted from HVSRs (Table 4). The S-wave velocity of 2.5 km/s found below the main interface is in good accordance with the shallow layers of the general 1D models for Alboran Sea obtained by Grevemeyer et al. [37] and Stich et al. [71] from arrival times of seismic phases of local earthquakes, which are respectively 2.4 and 2.8 km/s for the uppermost 3 km. The $V_P / V_S$ ratio of 1.76 for the halfspace is also consistent with their models which show values in the range 1.65 to 1.8. Simple 1D analysis predicts S-waves amplifications due to the sedimentary structure of up to 3 times at the resonance frequency (Fig. 17e-f).

As far as the authors' knowledge, our models provide the first bounds for $V_S$ of the whole stratigraphy for the sedimentary fillings in Campo de Dalías basin down to the Triassic basement by means of seismic techniques sensitive to this magnitude. The models coming from inversion of ambient noise measurements at deep boreholes in the outskirts of the town show Quaternary conglomerates with $V_S$ ranging from 820 to 1000 m/s, well above the surface velocity of most of the Quaternary units logged in the urban area. This result highlights the benefits of spending time in techniques capable of resolving weak velocity variations, such as the MASW and the SPAC method (Figs. 7 and 15). Pliocene units are represented in those models by calcarenites varying from 590 to 680 m/s and marls from 990 to 1200 m/s. Despite the aforementioned lack of local relationships between $N_{SPT}$ and $V_S$ for Pliocene units, this interval includes the value given from geotechnical analysis, previously considered as a rough approximation. Finally, layers between 1300 and 1600 m/s and those above 1800 m/s are identified as Tortonian calcarenites and Triassic dolomites, respectively. These older units had been only described in terms of $V_P$ and densities. The use of these $V_S$ models will help to perform significant simulations of the seismic response of the basin, which is the topic of an ongoing research work.

**Acknowledgements**

This article is dedicated to the memory of Antonio Sánchez Picón. The authors would like to thank J. Francisco Navarro López, Esther Martín Funes and Marina Arrien for helping in the field measurements. We also thank Luis Molina Sánchez for providing valuable borehole data. The authors gratefully acknowledge the support provided by the staff of the Municipal Archive and Local Police of El Ejido. The constructive comments of the editor and two anonymous referees significantly improved this article.

Funding: This work was supported by the Spanish Ministry of Economy and Competitiveness and the European Regional Development Fund [project number CGL2014-59908].